\documentclass[showpacs,twocolumn,aps,pra]{revtex4-1}

\usepackage{amssymb}
\usepackage{amsmath}
\usepackage{epsfig}
\usepackage[caption=false]{subfig}

\usepackage{morefloats}
\usepackage{hyperref}

\begin{document}

\title{The Wigner Current for Open Quantum Systems}

\author{William F. Braasch, Jr., Oscar D. Friedman, Alexander J. Rimberg, and Miles P. Blencowe}
\affiliation{Department of Physics and Astronomy, Dartmouth College, New Hampshire 03755, USA}

\date{\today}

\begin{abstract}
We extend the Wigner current vector field (Wigner current) construct to single bosonic mode quantum systems interacting with an environment. In terms of the Wigner function quasiprobability density and associated Wigner current, the open system quantum dynamics can be concisely expressed as a continuity equation. Through the consideration of the harmonic oscillator and additively driven Duffing oscillator in the bistable regime as illustrative system examples, we show how the evolving Wigner current vector field on the system phase space yields useful geometric insights concerning how quantum states decohere away due to interactions with the environment, as well as how they may be stabilized through the counteracting effects of the system anharmonicity (i.e., nonlinearity).         
\end{abstract}

\maketitle

\section{Introduction}
\label{sec:introduction}
The Wigner function~\cite{wigner32,hillery84,case08,curtright14} provides a particularly useful geometric representation of the state of a bosonic single mode quantum system as a real valued function on the two-dimensional system phase space that can be interpreted as a quasiprobability density. In particular, integrating the Wigner function with respect to the system position coordinate $x$ gives the marginal probability density in the momentum coordinate $p$ (and vice versa)
In terms of the Wigner function, the quantum expectation value of a (Weyl ordered) observable $A(x,p)$ is evaluated in exactly the same way as for the corresponding classical system described by a phase space probability distribution function. Furthermore, the master equation that describes the quantum dynamics of a bosonic single mode system interacting with its thermal environment gets mapped to a partial differential equation for the Wigner function dynamics that closely resembles the Fokker-Planck equation for the corresponding classical system statistical dynamics; the Wigner function dynamical equation may aptly be termed the `quantum Fokker-Planck' equation.

The Wigner function can take negative values and so is not a true probability distribution, however. The presence of regions in phase space where the Wigner function is negative is conventionally interpreted as a signature of nonclassicality in the quantum state; with the exception of Gaussian (i.e., coherent and squeezed) states, {\it all} pure states have negative-valued Wigner function regions and hence are nonclassical~\cite{hudson74}. Well known examples are the harmonic oscillator energy eigenstates or Fock states and so-called Schr\"{o}dinger cat states involving superpositions of different coherent states. In the presence of a thermal environment, such nonclassical, pure states will typically evolve into mixed states with Wigner function representations that are everywhere positive~\cite{zurek91}.

Given the close resemblance between the Wigner function representation of the open quantum system dynamical equations and the corresponding classical statistical dynamical equations, the Wigner function  has helped provide an understanding of how classical dynamics arises by approximation from the underlying quantum dynamics~\cite{zurek94,kohler98,habib98,monteoliva01,habib02,everitt05,dykman07,greenbaum07,katz08,stobinska08}. Of particular interest in this respect are nonlinear single mode systems such as the paradigmatic, driven damped Duffing oscillator. A number of  investigations  have employed the Wigner function representation to explore the resulting quantum phase space dynamics in parameter regimes where the corresponding classical nonlinear dynamics exhibits, for example, bistability or chaos~\cite{habib98,habib02,dykman07,greenbaum07,katz08}. By varying the system damping and noise (diffusion) due to coupling to the environment, the quantum to classical transition can be explored in a controllable and geometrically direct way by comparing the corresponding quantum Wigner function phase space and classical phase space pictures. 

In the present work we will extend the Wigner formulation of open, single mode bosonic system quantum dynamics  and take into account also the so-called Wigner current vector field (or `Wigner current' in short) on phase space~\cite{bauke11,steuernagel13}. We will show that the Wigner current allows a particularly concise reformulation of the quantum Fokker-Planck equation as a standard continuity equation, equating (via the familiar Gauss's theorem of vector calculus) the rate of change of the net Wigner quasiprobability  within some two dimensional region of phase space to the net Wigner current normal to the boundary enclosing the region.

Our key motivation for introducing the Wigner current and casting the bosonic mode system quantum dynamics into a continuity equation on phase space is to apply a more geometric approach to addressing the sought-after goal to generate macroscopic quantum states that are stable over long times against the decohering effects of the environment. By `macroscopic', we mean that the averaged number of energy quanta (e.g., photons, phonons etc.) is large in the stabilized, bosonic mode state, while by `quantum' we mean that the Wigner function representation of the state has significant negative regions in the system phase space. How large can a negative Wigner valued region be? Two classic theorems that can be easily generalized to mixed states establish that the Wigner function is generally bounded in magnitude by $(\pi\hbar)^{-1}=2/h$~\cite{baker58}, while the area of a given region where the Wigner function is negative can exceed $\hbar$, but where at least one of its normalized linear coordinate dimensions $x \sqrt{m\omega_0}$, $p/\sqrt{m\omega_0}$ must be of order $\sqrt{\hbar}$ or smaller~\cite{cartwright76}, with $m$  the effective mass and $\omega_0$ a characteristic harmonic oscillation frequency of the single mode bosonic system (see Sec. \ref{sec:duffing} for further details on the rescaled phase coordinate definitions).

Stabilized macroscopic quantum states  are useful not only for quantum information processing applications, but also for fundamental explorations, especially concerning how macroscopic a quantum state can be in the presence of unavoidable decohering environments.  As with the above-mentioned quantum-classical correspondence investigations, the Wigner function has served as a useful tool in addressing the generation and detection of nonclassical states of bosonic, single as well as few mode quantum systems. A range of investigations have been carried out involving optical~\cite{bimbard10,yoshikawa13}, microwave cavity~\cite{deleglise08} and superconducting circuit systems~\cite{hofheinz09,wang09,mallet11,eichler11,eichler12,shalibo12,shalibo13,kirchmair13,vlastakis13,wang16}, as well as nanomechanical systems~\cite{katz08,rips12,nation13,rips14,vanner15,abdi16}. 

Approaches to stabilizing quantum states involve measurement feedback to control the quantum system dynamics~\cite{joana16}, as well as so-called autonomous methods that do not require measurement feedback control. The latter typically involve `reservoir engineering', where the effective system-environment interaction is tailored in such a way as to evolve the system into a quantum state as well as to protect the state from the decohering effects of the environment~\cite{poyatos96,sarlette11, rips12,lin13,shankar13,Roy15,leghtas15,touzard18}.

Another approach to autonomously generating quantum states exploits the nonlinearities in the closed bosonic mode system dynamics---equivalently anharmonicities in the system Hamiltonian. The presence of anharmonicities can cause initial Gaussian states with associated positive Wigner functions to evolve into nonclassical states with associated negative valued Wigner functions (see, e.g., Ref.~\onlinecite{katz08}). In terms of the quantum Fokker-Planck dynamical equations for the Wigner function, the root cause of such evolution is the presence of a third or higher order position derivative term involving the system potential energy. Only when the potential energy is anharmonic is this term present and without this term, the Wigner function dynamical equation coincides with the classical Fokker-Planck equation. 

For example, in the case of the driven, damped Duffing oscillator with $ x^4$ anharmonicity and in the regime of bistable large and small amplitude oscillatory solutions for the classical dynamics, an initial coherent state will transiently evolve into a Schr\"{o}dinger cat-like state where the Wigner function displays a sequence of alternating negative and positive regions in between the corresponding large and small amplitude positive Wigner function peaks~\cite{katz08}. And in a classically chaotic regime, an initial coherent state will spread out in phase space, exhibiting a complex interference pattern of positive and sub-$\hbar$ (i.e., sub-Planckian) scale negative Wigner function regions~\cite{habib98}. However, depending on the environment temperature, such non-classical features will typically diffuse away for the usual device system-environment couplings, leaving a long time steady state that is closely approximated by the corresponding classical system Fokker-Planck equation.

Nevertheless, the question is still largely unresolved as to whether it might be possible to stabilize quantum states of a single mode bosonic system largely through its anharmonicities alone. In particular, for certain anharmonicity types and drives (whether externally or internally generated by the system dynamics), we may be able to prepare and maintain quantum states with significant associated negative Wigner function regions, despite the counteracting decoherence effects of environmental noise. Recent relevant developments in superconducting  microwave resonator  (as well as coupled nanomechanical resonator) circuits involving embedded Josephson junction elements provide strong motivation for pursuing this question~\cite{chen11,blencowe12,armour13,gramich13,chen14,rimberg14,armour15,souquet16,dambach16};  the Josephson junctions can induce strong effective anharmonicities in the microwave mode Hamiltonian, as well as internally generated drive tones through the ac-Josephson effect. One consequence is lasing-like behavior~\cite{chen14}, with the continuous, stimulated emission of amplitude-squeezed microwaves with large average photon number~\cite{armour13}.

The potential advantage of bringing the Wigner current  into play is that it can give a graphic geometric representation of how non-classical states form through the system Hamiltonian anharmonicity, as well as diffuse away due to the environment. By exploring the relative contributions to the net Wigner current  across the boundary of a given negative region that arise from the system Hamiltonian anharmonicity and from the interactions with the environment, we may be able to improve our understanding of how to `engineer' system Hamiltonian anharmonicities and drive tones so as to stabilize macroscopic bosonic quantum states in the presence of environmental noise. As an application of the geometric Wigner current construct for open quantum systems, the present work gives some initial steps in this direction.

Note that in the present work we do not attempt to address the largely open question as to how the negative Wigner function regions form in the first place; rather, we suppose that negative regions have already formed, and consider how the regions may be stabilized in the presence of environmental noise. Some promising first steps towards understanding how negative regions form from a Wigner current perspective are given in Refs. \onlinecite{oliva2017,oliva2018}.

In Sec.~\ref{sec:wigner}, we reformulate the quantum Fokker-Planck equation for a one dimensional anharmonic particle system interacting with a thermal bath as a continuity equation in terms of the Wigner function and associated current. In Sec.~\ref{sec:duffing} we consider as specific system examples the harmonic oscillator and additively driven Duffing oscillator, solving numerically for their Wigner functions and currents.  In Sec.~\ref{sec:discussion}, we take some first steps towards a geometric understanding from a Wigner current perspective concerning how nonclassical states may be stabilized. Concluding remarks are provided in Sec.~\ref{sec:conclusion}.

\section{Quantum Fokker-Planck Equation as a Continuity Equation}
\label{sec:wigner}
For a one-dimensional, mass $m$ particle with Hamiltonian $H=p^2/(2 m)+V(x,t)$, where $V(x,t)$ is the (time dependent)  potential energy,   a suitable Lindblad master equation that describes the quantum dynamics of the system state characterized by density matrix $\rho(t)$ interacting with an oscillator bath can be written as follows:
\begin{eqnarray}
\frac{d \rho}{dt}&=&-\frac{i}{\hbar}[H,\rho]+\frac{\gamma}{2}(\bar{n}+1)\left(2 a\rho a^{\dagger}-a^{\dagger}a\rho-\rho a^{\dagger}a\right)\cr
&&+\frac{\gamma}{2}\bar{n}\left(2a^{\dagger}\rho a-a a^{\dagger}\rho-\rho a a^{\dagger}\right),
\label{mastereq}
\end{eqnarray}
where $\gamma$ is the system energy damping rate, $\bar{n}=(e^{\hbar\omega_0/(k_BT)}-1)^{-1}$ is the Bose-Einstein thermal average occupation number of the temperature $T$ bath at the characteristic harmonic oscillation frequency $\omega_0$ of the system Hamiltonian. Strictly speaking, the master equation~(\ref{mastereq}) is valid to a good approximation  provided the system-environment interaction is weak: $\gamma\ll\omega_0$, the temperature is in the range $\hbar\gamma\ll k_B T\ll\hbar\omega_0$, and the anharmonic potential contribution $V-m\omega_0^2 x^2/2$ is sufficiently weak~\cite{haake86}. However, following frequent practice, we will assume that the master equation can still give reasonable open system quantum dynamics even when these conditions are not strictly adhered to. 

The Wigner function representation of the quantum state $\rho(t)$ as a real-valued function on phase space is defined as~\cite{wigner32,hillery84,case08,curtright14}
\begin{eqnarray}
W(x,p,t)&=&\frac{1}{\pi\hbar}\int_{-\infty}^{+\infty} dy\, e^{-2ipy/\hbar}\langle x+y|{\rho}(t)|x-y\rangle\cr
&=&\frac{1}{\pi\hbar}\int_{-\infty}^{+\infty} dp'\, e^{+2ip' x/\hbar}\langle p+p'|{\rho}(t)|p-p'\rangle.\cr
&&\label{wfeq}
\end{eqnarray}
Expressing the master equation~(\ref{mastereq}) in terms of the Wigner function (\ref{wfeq}), we obtain the so-called `quantum Fokker-Planck' equation
\begin{eqnarray}
&&\frac{\partial W}{\partial t}=-\frac{p}{m}\frac{\partial W}{\partial x}+\frac{\partial V}{\partial x}\frac{\partial W}{\partial p}\cr
&&+\sum_{n\geq 1}\frac{(-1)^n(\hbar/2)^{2n}}{(2n+1)!}\frac{\partial^{2n+1}}{\partial x^{2n+1}}V\frac{\partial^{2n+1}}{\partial p^{2n+1}} W\cr
&&+\frac{\gamma}{2}\frac{\partial}{\partial x}\left[x W+\hbar\left(\bar{n}+\frac{1}{2}\right)\frac{1}{m\omega_0}\frac{\partial W}{\partial x}\right]\cr
&&+\frac{\gamma}{2}\frac{\partial}{\partial p}\left[p W+\hbar\left(\bar{n}+\frac{1}{2}\right){m\omega_0}\frac{\partial W}{\partial p}\right].
\label{wfmastereq}
\end{eqnarray}

The Wigner current vector fields for the system~(\cite{bauke11,steuernagel13}) and environment are defined respectively as follows:
\begin{eqnarray}
{\mathbf{J}}_{\mathrm{sys}}=
\left( \begin{array}{c}
\frac{p}{m} W\\
-\sum_{n=0} \frac{(-1)^{n}(\hbar/2)^{2n}}{(2n+1)!}\partial_x^{(2n+1)}V\partial_p^{(2n)}W \\
\end{array} \right)
\label{sysfloweq}
\end{eqnarray}
and
\begin{eqnarray}
{\mathbf{J}}_{\mathrm{env}}=
-\frac{\gamma}{2}\left( \begin{array}{c}
x W+\hbar\left(\bar{n}+\frac{1}{2}\right)(m\omega_0)^{-1}\partial_x W\\
p W+\hbar\left(\bar{n}+\frac{1}{2}\right) m\omega_0 \partial_p W \\
\end{array} \right),
\label{envfloweq}
\end{eqnarray}
where the first row is the position $x$ component  and the second row is momentum $p$ component of the current vector, and where we have used the shorthand notation: $\partial x \equiv\frac{\partial}{\partial x}$, and $\partial p \equiv\frac{\partial}{\partial p}$. 
The environment current can be further decomposed as a sum of damping and diffusion contributions: ${\mathbf{J}}_{\mathrm{env}}={\mathbf{J}}_{\mathrm{damp}}+{\mathbf{J}}_{\mathrm{diff}}$, where
\begin{eqnarray}
{\mathbf{J}}_{\mathrm{damp}}=
-\frac{\gamma}{2}\left( \begin{array}{c}
x W\\
p W\\
\end{array} \right)
\label{dampfloweq}
\end{eqnarray}
and
\begin{eqnarray}
{\mathbf{J}}_{\mathrm{diff}}=
-\frac{\gamma\hbar}{2}\left(\bar{n}+\frac{1}{2}\right)\left( \begin{array}{c}
(m\omega_0)^{-1}\partial_x W\\
m\omega_0 \partial_p W \\
\end{array} \right).
\label{difffloweq}
\end{eqnarray}
In terms of the system and environment currents, the master equation for the Wigner function~(\ref{wfmastereq}) takes the concise form of a continuity equation:
\begin{equation}
\frac{\partial W}{\partial t} +\nabla\cdot {\mathbf{J}}=0,
\label{continuityeq}
\end{equation}
where ${\mathbf{J}}={\mathbf{J}}_{\mathrm{sys}}+{\mathbf{J}}_{\mathrm{env}}$ and $\nabla=(\partial_x,\partial_p)$.

\section{Harmonic and Duffing Oscillator Wigner Currents}
\label{sec:duffing}
The driven Duffing oscillator is characterized by the anharmonic $+$ additive  driving potential
\begin{equation}
V(x,t)=\frac{1}{2}m\omega_0^2 x^2+\frac{\lambda}{4} x^4 -x F\cos(\omega_d t),
\label{duffingpoteq}
\end{equation}
where the parameter $\lambda$ gives the strength of the anharmonic potential, the parameter $F$ gives the strength of the time-dependent sinusoidal drive, and $\omega_d$ is the drive frequency. Substituting Eq.~(\ref{duffingpoteq}) into  Eq.~(\ref{sysfloweq}), we obtain for the driven Duffing oscillator system Wigner current:
\begin{eqnarray}
{\mathbf{J}}_{\mathrm{Duff}}=
\left( \begin{array}{c}
\frac{p}{m} W\\
\left[-m\omega^2_0 x+F\cos(\omega_d t)-\lambda x^3+\frac{\hbar^2\lambda }{4} x \partial_p^2\right] W \\
\end{array} \right).\cr
&&\label{duffingsysfloweq}
\end{eqnarray}
For the harmonic oscillator the system Wigner current simplifies to 
\begin{eqnarray}
{\mathbf{J}}_{\mathrm{HO}}=
\left( \begin{array}{c}
\frac{p}{m} W\\
-m\omega^2_0 x W \\
\end{array} \right).
\label{hosysfloweq}
\end{eqnarray}

It is convenient to work in terms of dimensionless forms of the Wigner function and current. In terms of the length unit $x_0=\sqrt{\hbar/(m\omega_0)}$ and time unit $t_0=\omega_0^{-1}$, we transform the various coordinates and parameters into dimensionless form as follows: $\tilde{x}=x/x_0$, $\tilde{p}=p/(m\omega_0 x_0)$, $\tilde{F}=x_0 F/(\hbar\omega_0)$, $\tilde{\lambda}=\lambda x_0^4/(\hbar\omega_0)$, $\tilde{\gamma}=\gamma/\omega_0$, $\tilde{\omega}_d=\omega_d/\omega_0$, and $\tilde{t}=\omega_0 t$, where the tilde denotes the dimensionless form. The dimensionless form for the Wigner function is
\begin{eqnarray}
\tilde{W}&=&\hbar W\cr
&=&\frac{1}{\pi}\int_{-\infty}^{+\infty} dy\, e^{-2ipy/\hbar}\langle x+y|{\rho}(t)|x-y\rangle\cr
&=&\frac{1}{\pi}\int_{-\infty}^{+\infty} d\tilde{y}\, e^{-2i\tilde{p}\tilde{y}}\langle \tilde{x}+\tilde{y}|{\rho}(t)|\tilde{x}-\tilde{y}\rangle,
\label{dimlesswfeq}
\end{eqnarray}
where $|\tilde{x}\rangle=\sqrt{x_0}|x\rangle$ [so that $\langle\tilde{x}|\tilde{x}'\rangle=\delta(\tilde{x}-\tilde{x}')$]. The continuity equation becomes in dimensionless form:
\begin{equation}
\frac{\partial \tilde{W}}{\partial \tilde{t}} +\tilde{\nabla}\cdot \tilde{{\mathbf{J}}}=0,
\label{dimlesscontinuityeq}
\end{equation}
where ${\tilde{\mathbf{J}}}=\tilde{{\mathbf{J}}}_{\mathrm{Duff}}+\tilde{{\mathbf{J}}}_{\mathrm{env}}$, with  
\begin{eqnarray}
\tilde{{\mathbf{J}}}_{\mathrm{Duff}}=
\left( \begin{array}{c}
\tilde{p} \tilde{W}\\
\left[-\tilde{x}+\tilde{F}\cos(\tilde{\omega}_d \tilde{t})-\tilde{\lambda} \tilde{x}^3+\frac{\tilde{\lambda}}{4} \tilde{x} \partial_{\tilde{p}}^2\right] \tilde{W} \\
\end{array} \right),\cr
&&\label{dimlessduffingsysfloweq}
\end{eqnarray}
and
\begin{equation}
\tilde{{\mathbf{J}}}_{\mathrm{env}}=\tilde{{\mathbf{J}}}_{\mathrm{damp}}+\tilde{{\mathbf{J}}}_{\mathrm{diff}},
\label{dimlessenvfloweq}
\end{equation}
with
\begin{eqnarray}
\tilde{{\mathbf{J}}}_{\mathrm{damp}}=
-\frac{\tilde{\gamma}}{2}\left( \begin{array}{c}
\tilde{x} \tilde{W}\\
\tilde{p}\tilde{W}\\
\end{array} \right)
\label{dimlessdampfloweq}
\end{eqnarray}
and
\begin{eqnarray}
\tilde{{\mathbf{J}}}_{\mathrm{diff}}=
-\frac{\tilde{\gamma}}{2}\left(\bar{n}+\frac{1}{2}\right)\left( \begin{array}{c}
\partial_{\tilde{x}} \tilde{W}\\
\partial_{\tilde{p}} \tilde{W} \\
\end{array} \right).
\label{dimlessdifffloweq}
\end{eqnarray}
For the harmonic oscillator, we have for the dimensionless current: $\tilde{\mathbf{J}}=\tilde{\mathbf{J}}_{\mathrm{HO}}+\tilde{\mathbf{J}}_{\mathrm{env}}$,
with
\begin{eqnarray}
\tilde{\mathbf{J}}_{\mathrm{HO}}=
\left( \begin{array}{c}
\tilde{p}\tilde{W}\\
-\tilde{x} \tilde{W} \\
\end{array} \right)
\label{dimlesshosysfloweq}
\end{eqnarray}
and $\tilde{\mathbf{J}}_{\mathrm{env}}$ given by Eq.~(\ref{dimlessenvfloweq})  
From now on, we drop the tildes for notational convenience, the dimensionless form of the parameters and coordinates understood.

\begin{figure*}[htp]
\centering
\subfloat[$t=0$ (initial state)]{\label{fig:figure1a}
  \includegraphics[width=0.32\textwidth]{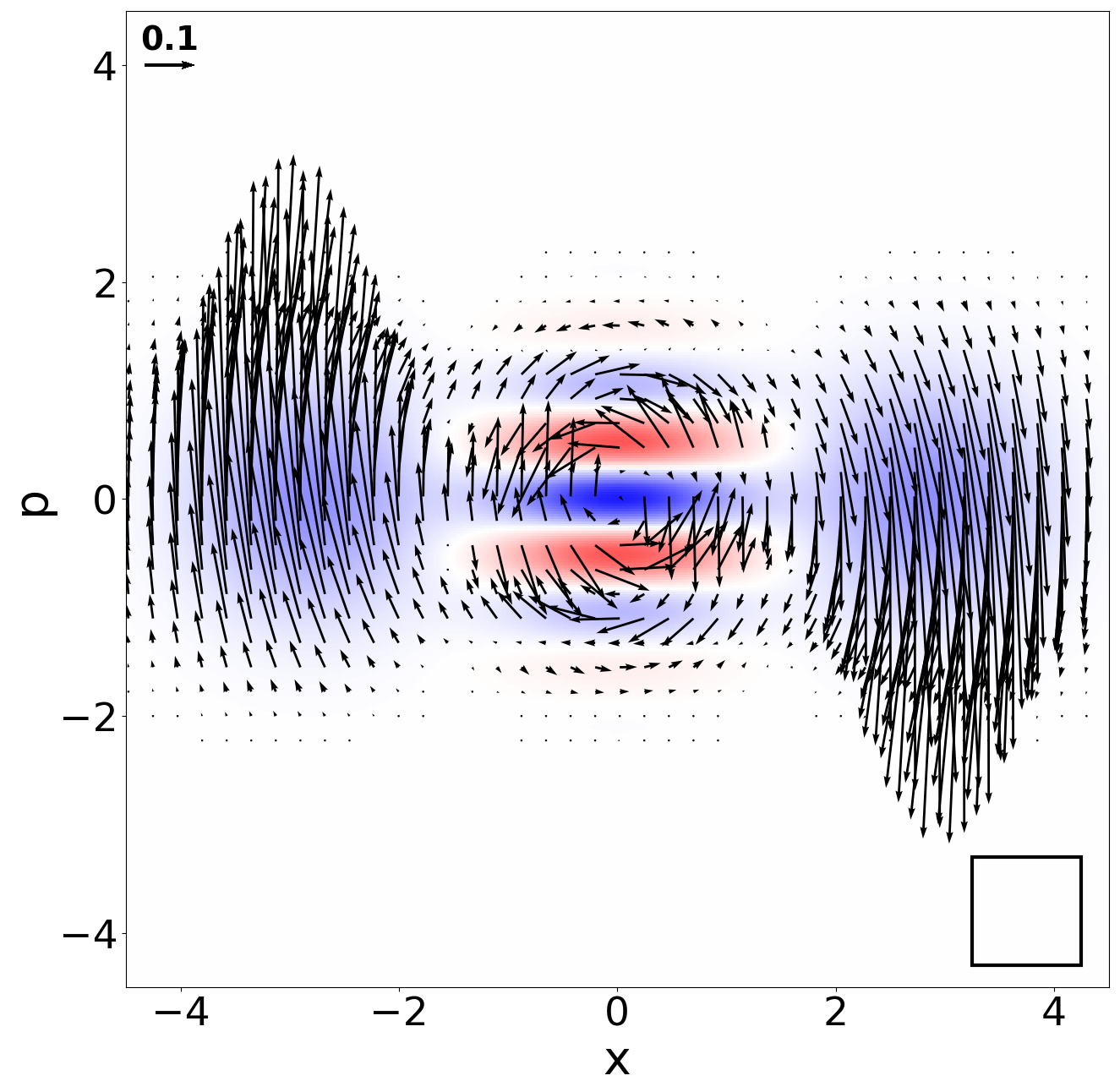}
}
\subfloat[$t=4\tau$]{\label{fig:figure1b}
  \includegraphics[width=0.32\textwidth]{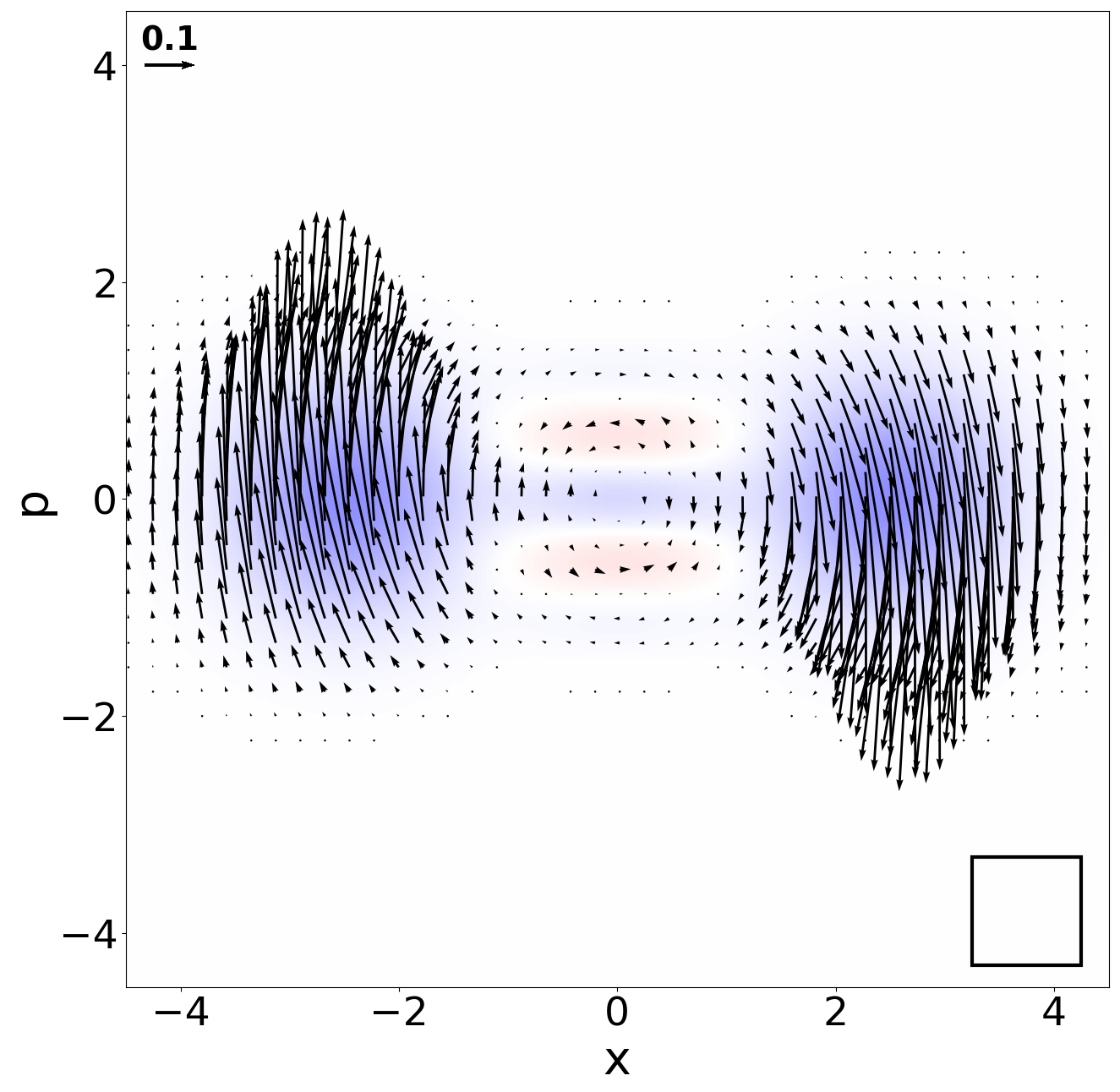}
}
\subfloat[$t=100\tau$]{\label{fig:figure1c}
  \includegraphics[width=0.32\textwidth]{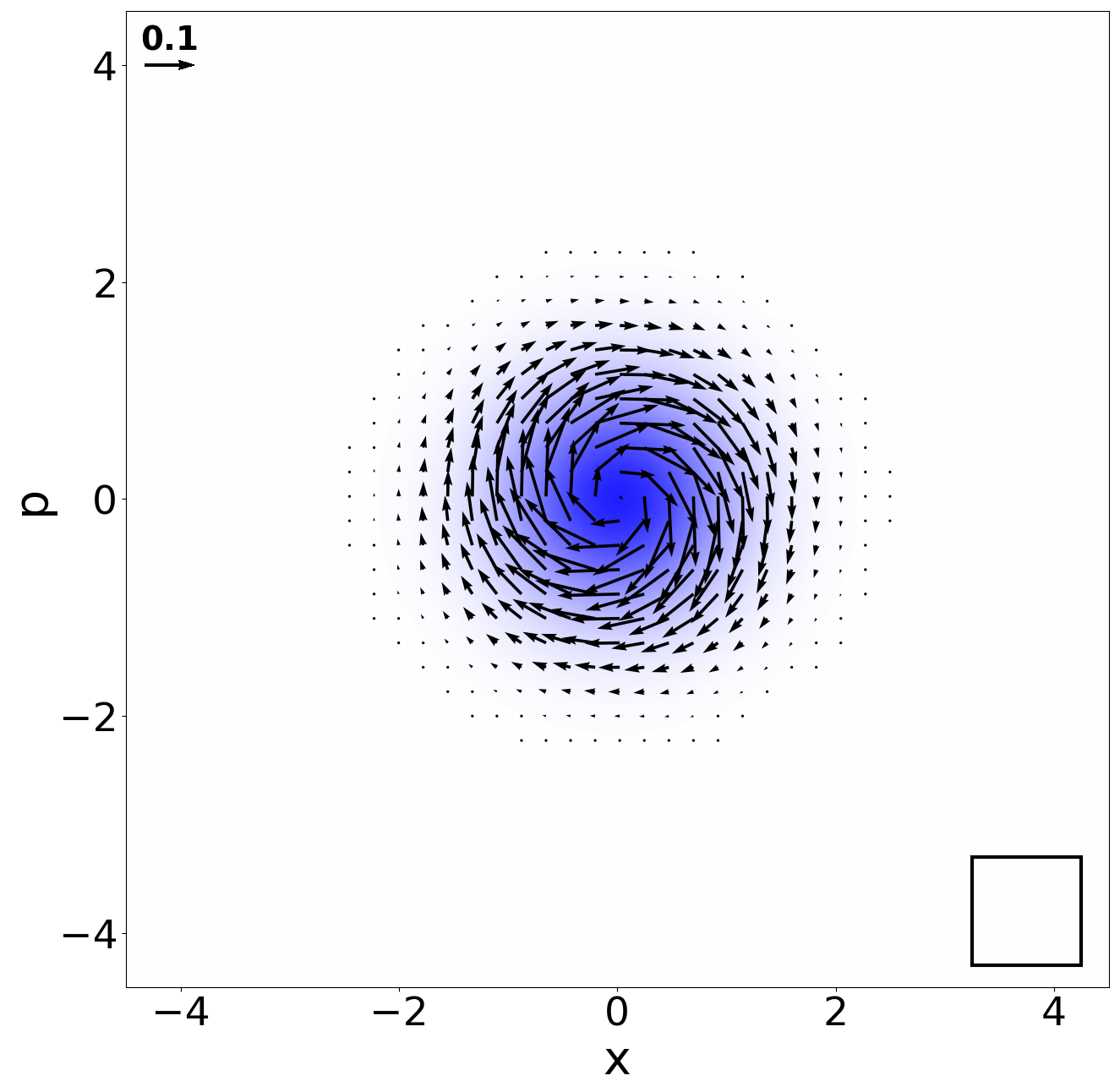}
}
\caption{Snapshots of evolving harmonic oscillator Wigner function and associated current vector field ${\mathbf{J}}={\mathbf{J}}_{\mathrm{HO}}+{\mathbf{J}}_{\mathrm{env}}$ for an initial  superposition of coherent states with separation $x=6$; the damping rate $\gamma=0.01$ and bath temperature $T=0$. }
\label{fig:figure1}
\end{figure*}

\begin{figure*}[htp]
\centering
\subfloat[$t=0$ (initial state)]{\label{fig:figure2a}
  \includegraphics[width=0.32\textwidth]{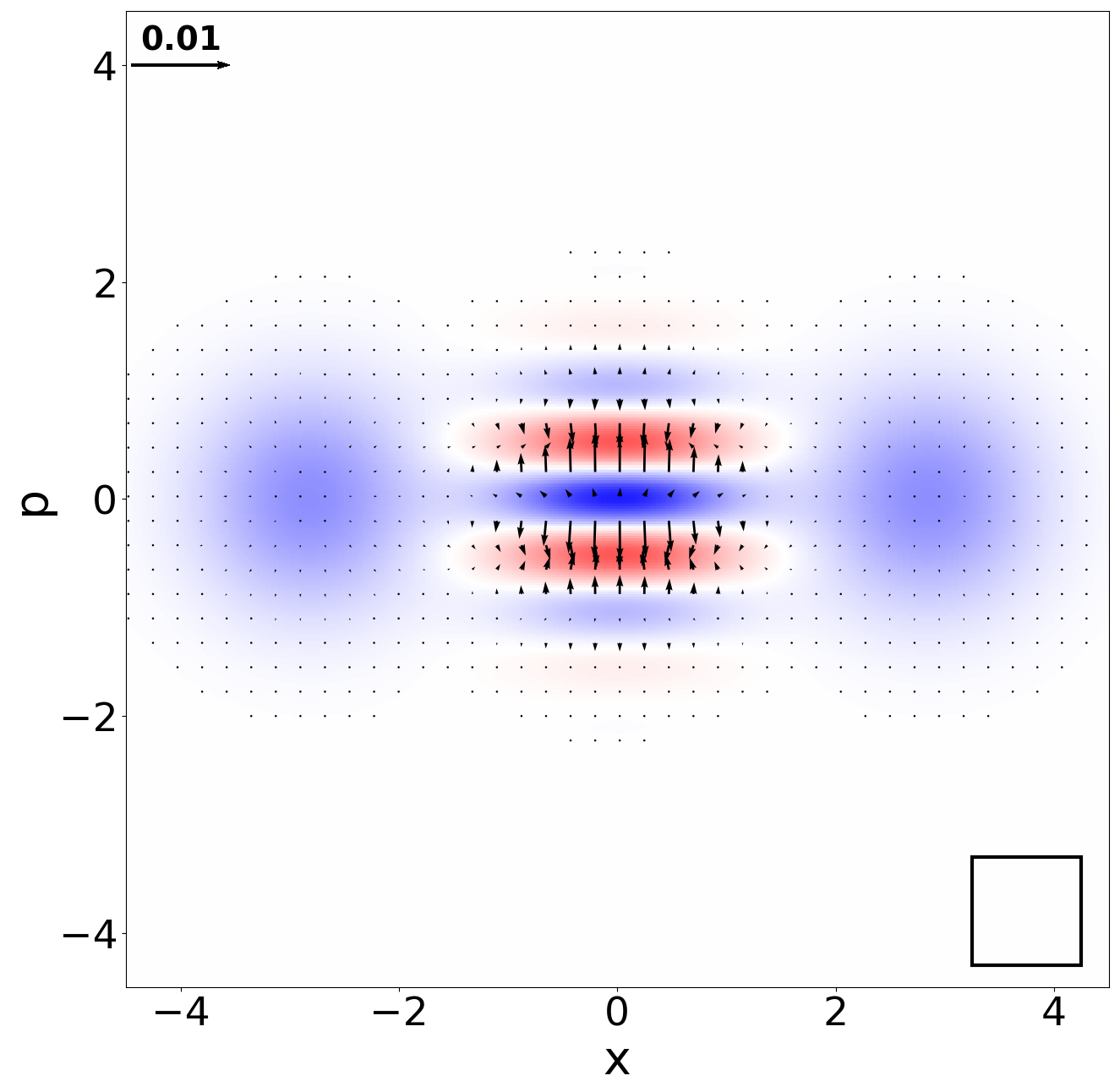}
}
\subfloat[$t=4\tau$]{\label{fig:figure2b}
  \includegraphics[width=0.32\textwidth]{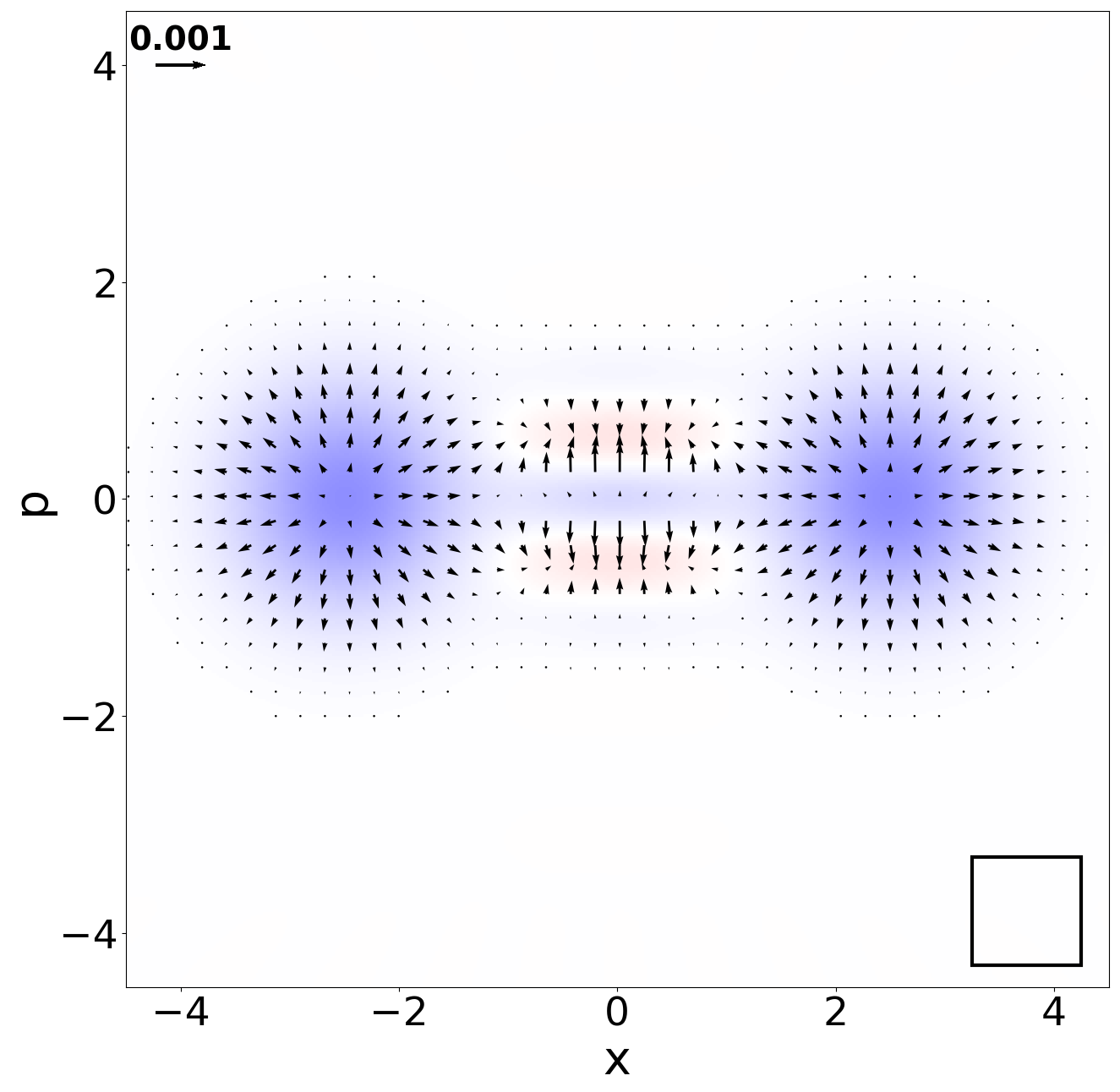}
}
\subfloat[$t=100\tau$]{\label{fig:figure2c}
  \includegraphics[width=0.32\textwidth]{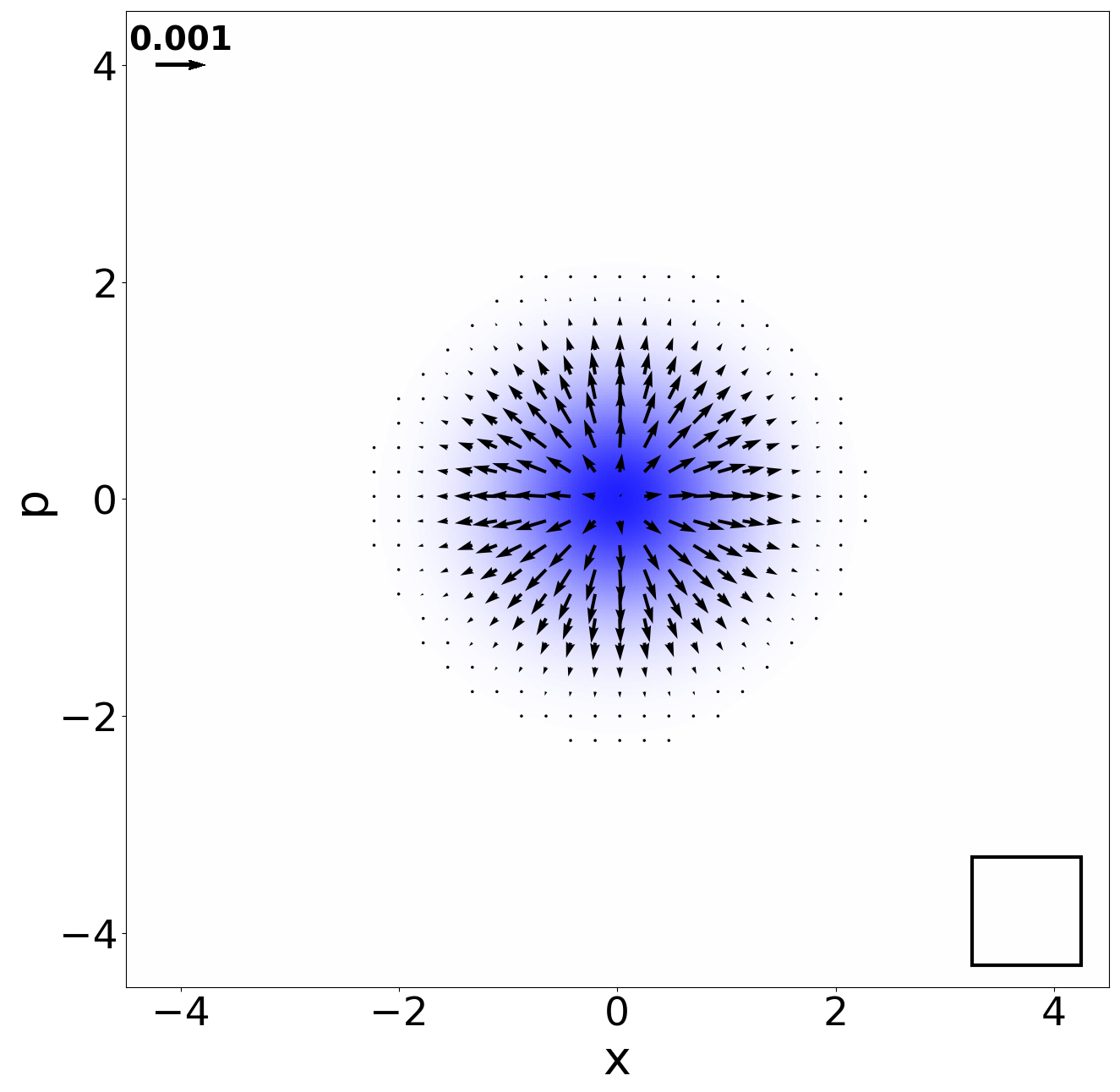}
}
\caption{Snapshots of evolving harmonic oscillator Wigner function and associated environmental diffusion current vector field ${\mathbf{J}}_{\mathrm{diff}}$ for an initial  superposition of coherent states with separation $x=6$; the damping rate $\gamma=0.01$ and bath temperature $T=0$.}
\label{fig:figure2}
\end{figure*}

\begin{figure*}[htp]
\centering
\subfloat[$t=0$ (initial state)]{\label{fig:figure3a}
  \includegraphics[width=0.32\textwidth]{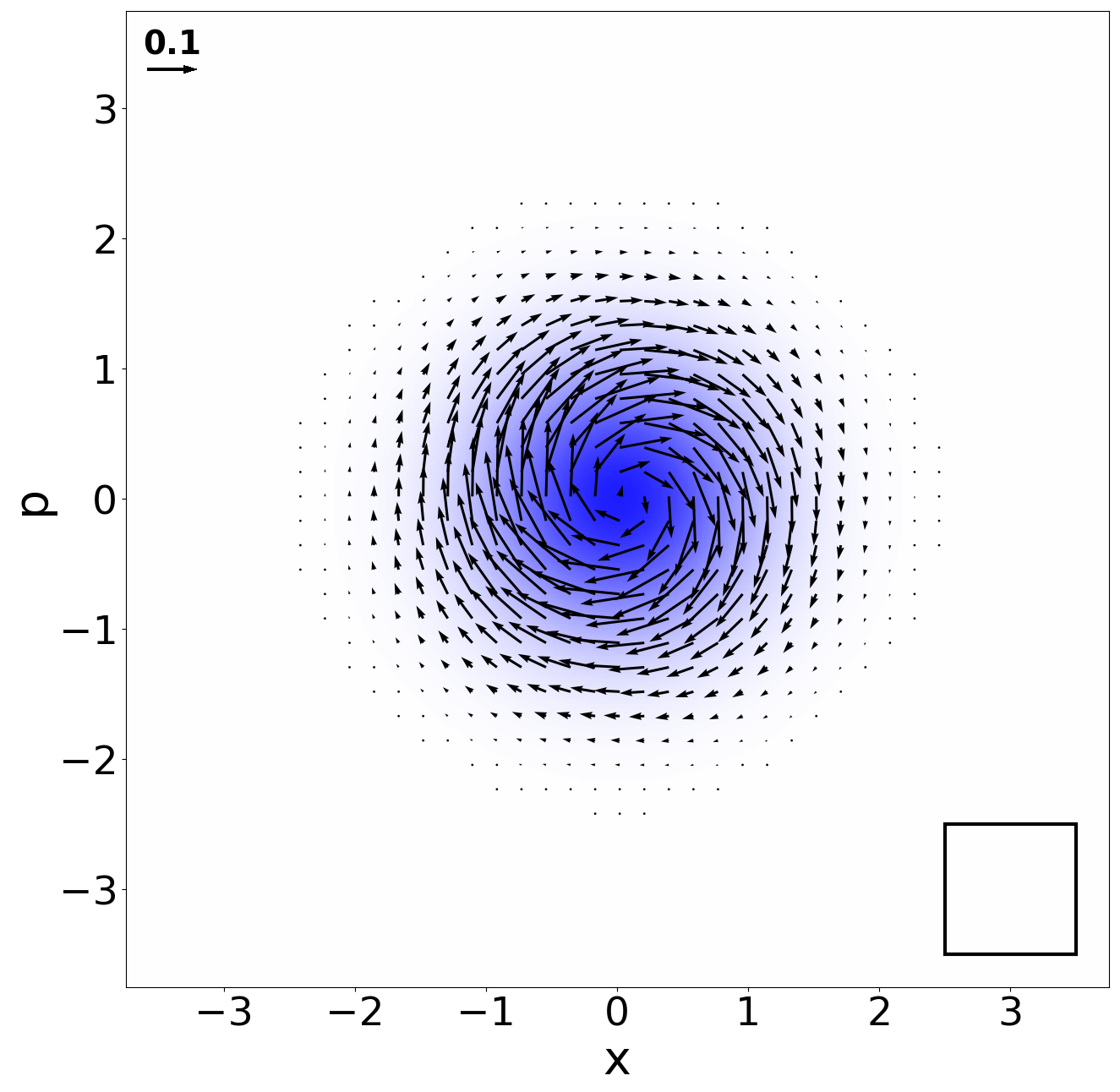}
}
\subfloat[$t=18\tau_d$]{\label{fig:figure3b}
  \includegraphics[width=0.32\textwidth]{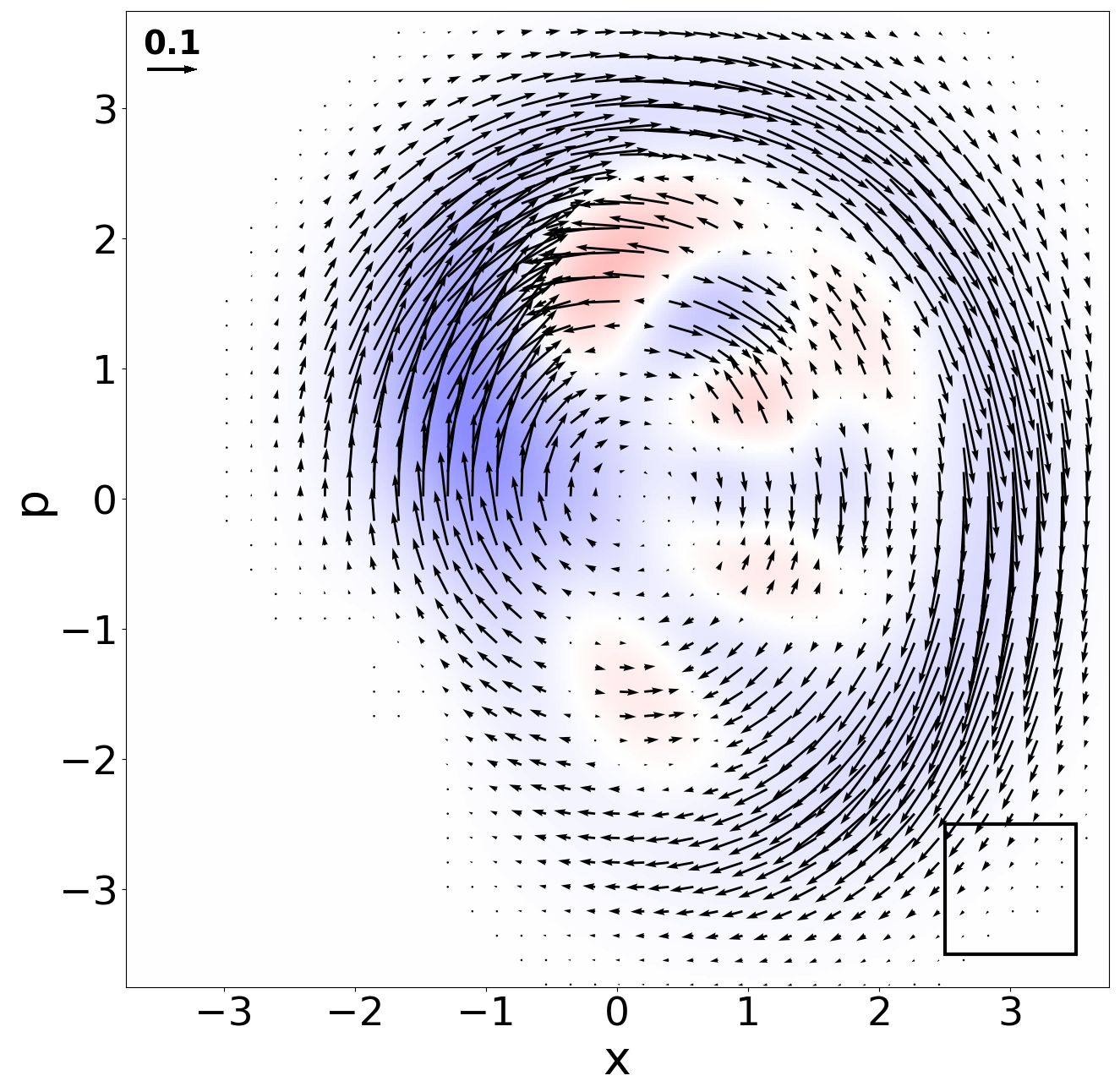}
}
\subfloat[$t=300\tau_d$]{\label{fig:figure3c}
  \includegraphics[width=0.32\textwidth]{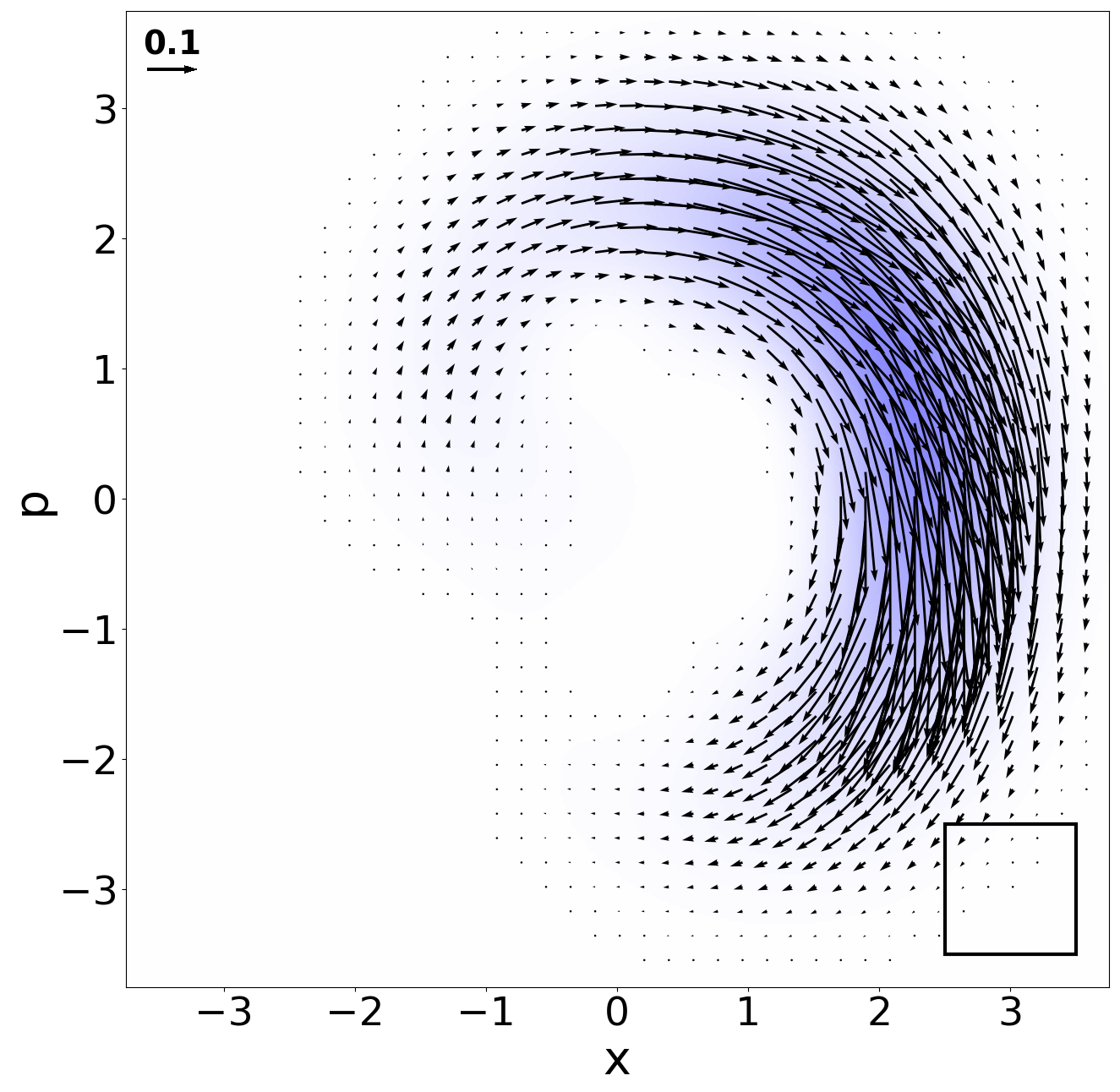}
}

\caption{Snapshots of evolving Duffing oscillator Wigner function and associated current vector field ${\mathbf{J}}={\mathbf{J}}_{\mathrm{Duff}}+{\mathbf{J}}_{\mathrm{env}}$ for an initial undisplaced coherent state; the damping rate $\gamma=0.01$ and bath temperature $T=0$.}
\label{fig:figure3}
\end{figure*}

\begin{figure*}[htp]
\centering
\subfloat[$t=0$ (initial state)]{\label{fig:figure4a}
  \includegraphics[width=0.32\textwidth]{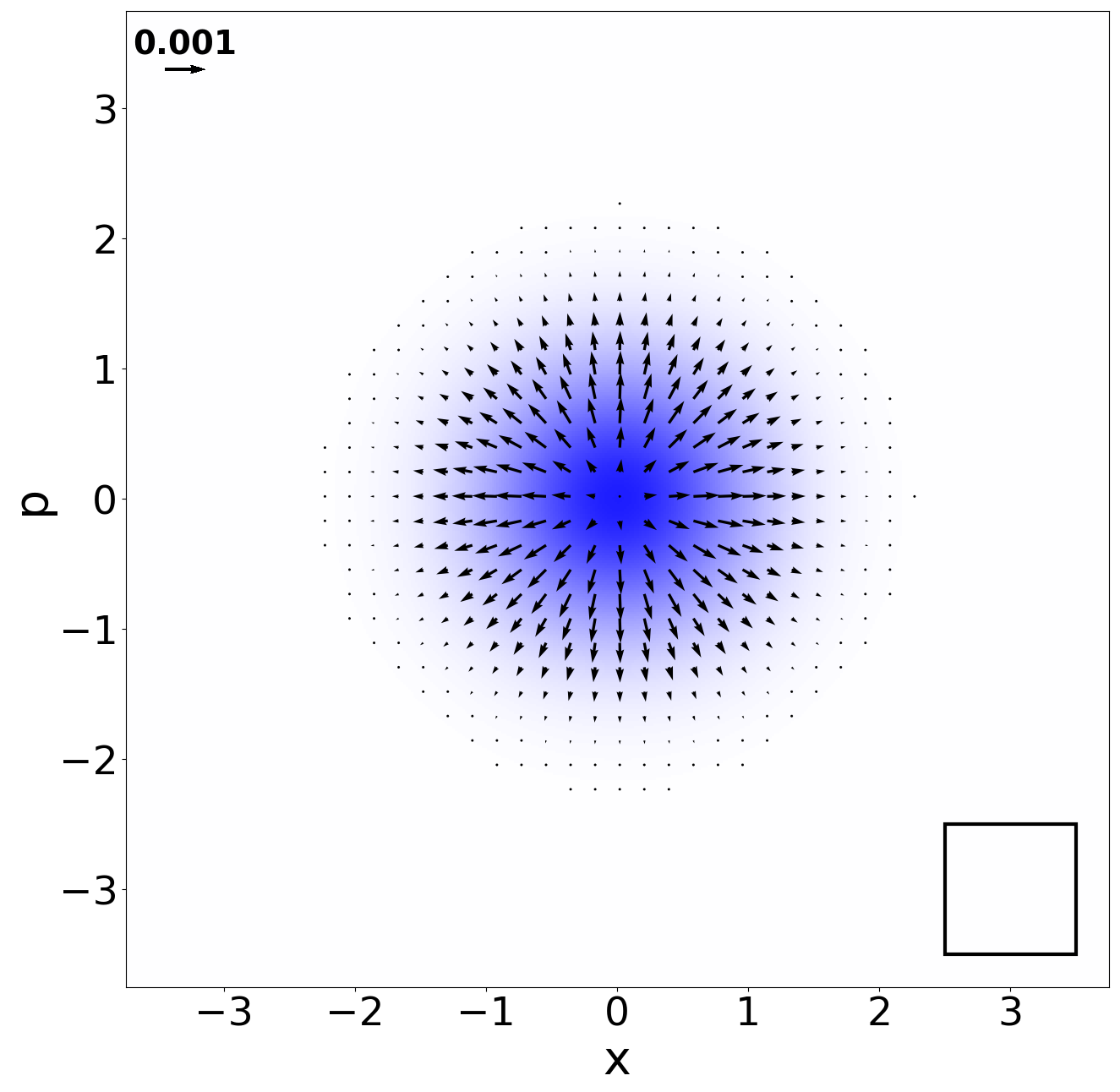}
}
\subfloat[$t=18\tau_d$]{\label{fig:figure4b}
  \includegraphics[width=0.32\textwidth]{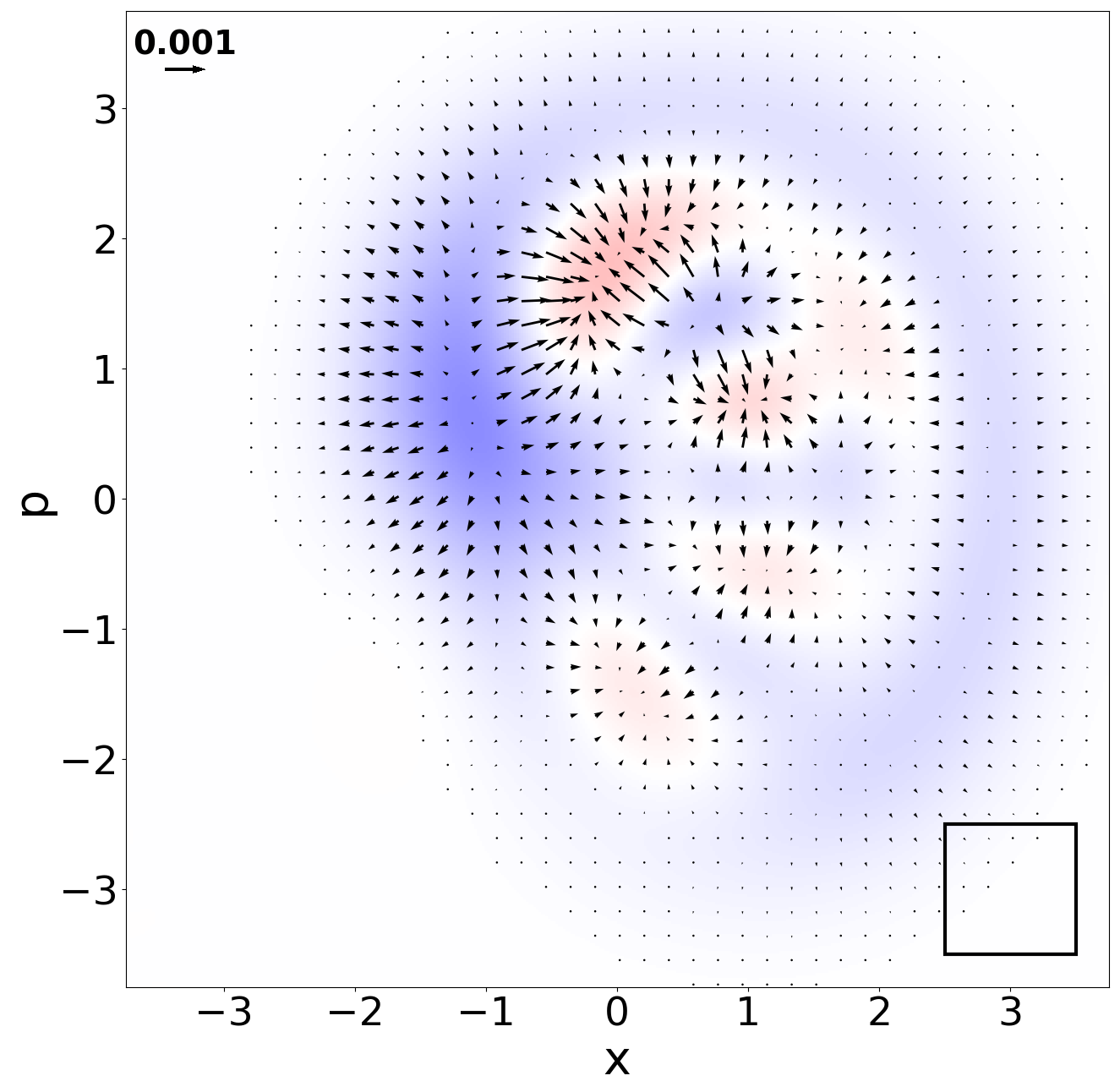}
}
\subfloat[$t=300\tau_d$]{\label{fig:figure4c}
  \includegraphics[width=0.32\textwidth]{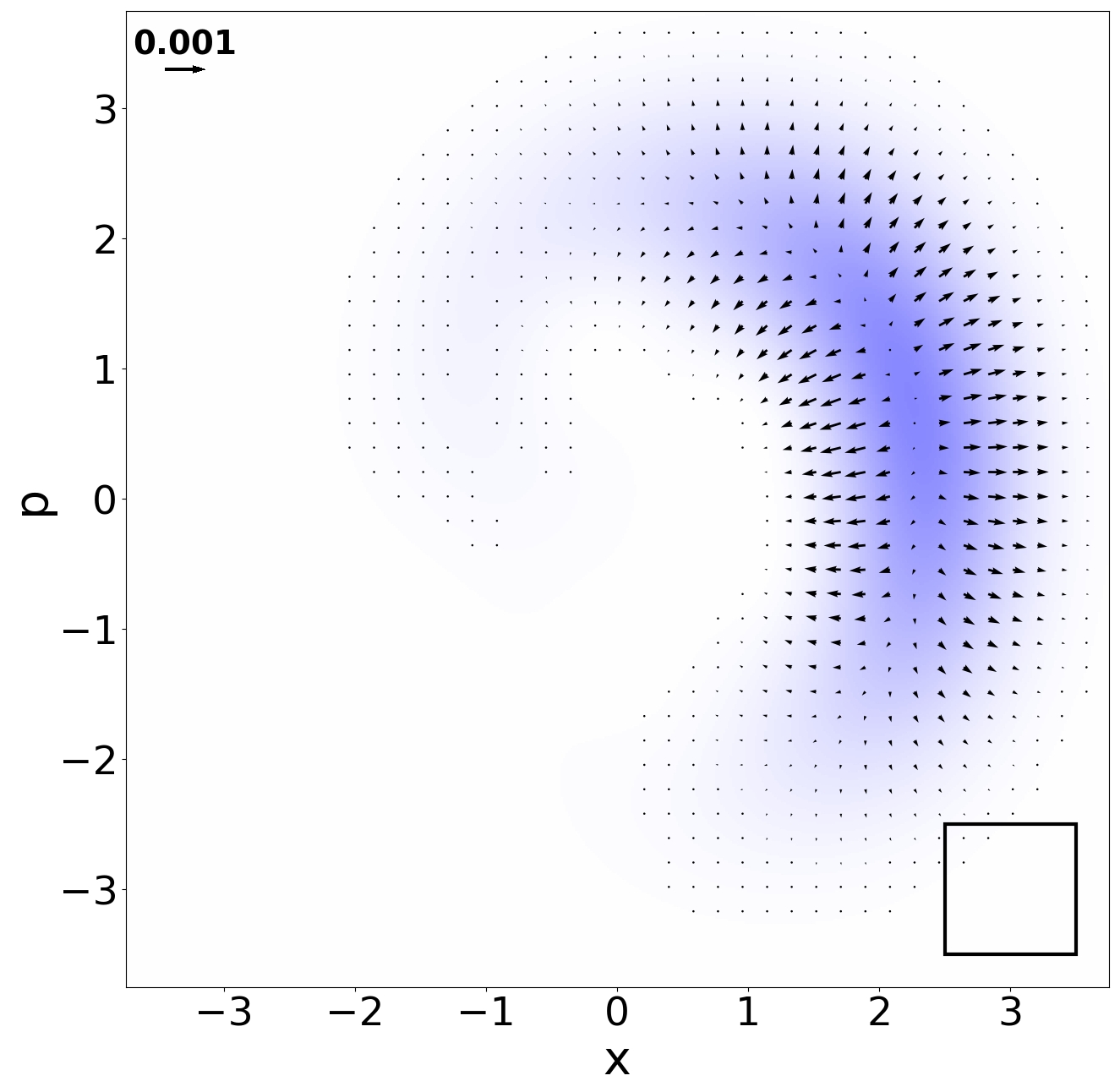}
}
\caption{Snapshots of evolving Duffing oscillator Wigner function and associated environmental diffusion current vector field ${\mathbf{J}}_{\mathrm{diff}}$ for an initial undisplaced coherent state; the damping rate $\gamma=0.01$ and bath temperature $T=0$.}
\label{fig:figure4}
\end{figure*}

In Figs.~\ref{fig:figure1}-\ref{fig:figure4}, we show example numerical solutions to the Wigner function $W$ and associated current vector field ${\mathbf{J}}$ for the undriven, open harmonic and driven Duffing oscillator systems. This involves first solving the Lindblad master equation~(\ref{mastereq}) for the system density matrix $\rho(t)$ using QuTiP~\cite{johansson13} and then evaluating the Wigner function and current in terms of the density matrix; the source code can be obtained from  Ref.~\cite{braaschfriedman19}.
Although the Wigner function time dependence for the open harmonic oscillator system can be determined analytically~\cite{kim92,paz93}, we nevertheless solve the harmonic oscillator master equation numerically as a check on the validity of our code. 

Figure~\ref{fig:figure1}  shows snapshots of the evolving Wigner function and associated current ${\mathbf{J}}={\mathbf{J}}_{\mathrm{HO}}+{\mathbf{J}}_{\mathrm{env}}$ for the harmonic oscillator initially in an initial superposition of coherent states separated by $x=6$; the snapshot times are given in multiples of the free oscillation period $\tau = 2\pi/\omega_0=2\pi t_0$. The damping rate is chosen to be $\gamma=0.01$ and the bath temperature is set to zero. Regions color-coded blue correspond to positive Wigner function value, red regions correspond to negative Wigner function value, while the local color density gives a measure of the Wigner function magnitude.   A unit area square corresponding to Planck's constant $\hbar$ in our dimensionless units is indicated at the bottom right of each figure to give the scale, while the arrow legend at the top left of each figure indicates the scale for the current vector field. Figure~\ref{fig:figure2} shows the same evolving Wigner function snapshots as in Fig.~\ref{fig:figure1} but with just the environmental diffusion current ${\mathbf{J}}_{\mathrm{diff}}$ (\ref{dimlessdifffloweq}) indicated. In the final indicated snapshots corresponding to $t=100\tau$ [Figs.~\ref{fig:figure1}-\ref{fig:figure2}(c)], the Wigner function and current hardly change between subsequent snapshots separated by a free oscillation period, 
indicating that the system dynamics has reached a steady state to a good approximation. This is to be expected given that $\gamma t=2 \pi$, i.e., the final snapshot time is approximately six times longer than the harmonic oscillator relaxation time.  

Figure~ \ref{fig:figure3} shows snapshots of the evolving Wigner function and associated current ${\mathbf{J}}={\mathbf{J}}_{\mathrm{Duff}}+{\mathbf{J}}_{\mathrm{env}}$ for the driven Duffing oscillator initially in an undisplaced coherent state; the snapshot times are given in multiples of the drive period $\tau_d = 2\pi/\omega_d$. We choose the dimensionless Duffing oscillator parameter values $\lambda=0.05$ (anharmonic strength), $\omega_d=1.09$ (drive frequency), $F=0.092$ (drive strength), and $\gamma=0.01$ (damping rate), with the bath temperature set to zero. These parameter values result in the classical Duffing oscillator exhibiting bistability for the steady state dynamics at zero temperature, corresponding to coexisting small and large amplitude oscillations. For the above parameter choices, these small and large steady state amplitudes are $0.52$ and $2.46$, respectively.  Figure~\ref{fig:figure4} shows the same evolving Wigner function snapshots as in Fig.~\ref{fig:figure3}, but with just the environmental diffusion current ${\mathbf{J}}_{\mathrm{diff}}$ (\ref{dimlessdifffloweq}) indicated. In the final indicated snapshots corresponding to $t=300\tau_d$ [Figs.~\ref{fig:figure3}-\ref{fig:figure4}(c)], the Wigner function and current hardly change between subsequent snapshots separated by a drive period. 
These final snapshots should therefore correspond pretty accurately to the long time limit steady state Wigner function and current.

\section{Discussion}
\label{sec:discussion}
Common to the harmonic and Duffing oscillator quantum dynamics indicated in Figs.~\ref{fig:figure1} and \ref{fig:figure3}, the direction of the current ${\mathbf{J}}$ in the regions of positive-valued Wigner function is clockwise about the phase space origin, just as is the case for an evolving classical probability density that results from solving the corresponding classical Fokker-Planck equation for some initial probability distribution; for the harmonic oscillator system, the Wigner current continuity equation~(\ref{continuityeq}) coincides with the classical, Brownian motion Fokker-Planck equation, while for the Duffing oscillator the Wigner current continuity equation~(\ref{continuityeq}) differs from the  classical Fokker-Planck equation only in the presence of the system quantum current term $(0,\lambda x\partial_p^2 W/4)$ [see Eq.~(\ref{dimlessduffingsysfloweq})]. 
In contrast, the current direction in the regions of negative-valued Wigner function is {\em{counterclockwise}}, i.e., in the opposite direction to the corresponding classical current~\cite{bauke11,steuernagel13,albarelli2016}.

The harmonic and time-dependent drive contributions to ${\mathbf{J}}_{\mathrm{Duff}}$ [Eq. (\ref{dimlessduffingsysfloweq})]  dominate the Wigner current. In principle, one can go to the rotating frame of the drive so that the nontrivial anharmonic and environment contributions dominate. However, the resulting algebraic expressions in the rotating frame become quite involved. While considerable simplification results if the rotating wave approximation (RWA) is made, we prefer not to go to the rotating frame and perform the RWA in the present work. As we shall see below, the dynamics throughout a given drive period of the non-vanishing quantum and diffusion contributions to the Wigner current on a negative region boundary  suggests a possible way to stabilize negative regions in the long time limit; such a dynamics would be effectively averaged over if the RWA were carried out.  

In Figs.~\ref{fig:figure2} and \ref{fig:figure4},  we can see that for any negative-valued Wigner function region, the diffusion contribution to the environmental current ${\mathbf{J}}_{\mathrm{diff}}$ is always directed inwards on the boundary of the negative region, with the result that the environmental diffusion current acts to destroy negative regions. This is just the process of decoherence viewed in terms of the Wigner current.    

In order to gain a better understanding of the Wigner function evolution for non-classical states, let us suppose that the Wigner function at some given time instant $t$ is negative in certain regions of phase space. This is the case for the initial coherent state superposition example considered above (see Figs.~\ref{fig:figure1} and \ref{fig:figure2}), while for the Duffing oscillator, we see that negative Wigner function regions are generated through the dynamics (Figs.~\ref{fig:figure3} and \ref{fig:figure4}). Consider a particular negative region with phase space area $A(t)$ and boundary $\partial A(t)$, where the indicated $t$-dependence accounts for the fact that the negative region evolves in time. In particular, the boundary is defined by  $\left. W(x,p,t)\right|_{\partial A(t)}=0$. A measure of the degree of negativity of the region is given by the negative `volume' under the integral
$\int_{A(t)} dx dp\, W(x,p,t)$. From Eqs.~(\ref{dimlesscontinuityeq})-(\ref{dimlessdifffloweq}) and Gauss's theorem, the time rate of change of this negative volume is
\begin{eqnarray}
&&\frac{d}{dt}\int_{A(t)} dx dp\, W(x,p,t)=-\frac{\lambda}{4} \int_{\partial A(t)} ds\, {\mathbf{n}}\cdot\left(0,x\right) \frac{\partial^2 W}{\partial p^2} \cr
&&+ \frac{\gamma}{2}\left(n+\frac{1}{2}\right) \int_{\partial A(t)}ds\, {\mathbf{n}}\cdot\nabla W,
\label{negduffrateeq}
\end{eqnarray}
where we have used the fact that the Wigner function vanishes on the boundary $\partial A(t)$, $s$ parametrizes the boundary curve, and ${\mathbf{n}}$ is the unit vector outwards normal to the curve. 

For the harmonic oscillator system, the first term on the right hand side of the equals sign in Eq.~(\ref{negduffrateeq}) vanishes (since $\lambda=0$) and the rate of change of the region negativity is affected solely by the environmental diffusion current~(\ref{dimlessdifffloweq}).  Since the Wigner function is by definition negative on the interior region and positive on at least the immediate exterior region of the boundary $\partial A(t)$, the gradient $\nabla W$ points outwards so that ${\mathbf{n}}\cdot \nabla W\geq 0$ everywhere on the boundary. Therefore, for the harmonic oscillator we have that 
\begin{equation}
\frac{d}{dt}\int_{A(t)} dx dp\, W(x,p,t)\geq 0 
\label{increasingWFeq}
\end{equation}
and we thus see that the size of a negative region always decreases with time at a rate governed by the environmental diffusion current. That the environment causes decoherence for a harmonic oscillator initially in a quantum superposition state is of course well-known. Nevertheless, in our view there is value in picturing the process of decoherence from a geometric, current perspective, especially  in the case where the oscillator is anharmonic.

\begin{figure*}[htp]
\centering
\includegraphics[width=0.24\textwidth]{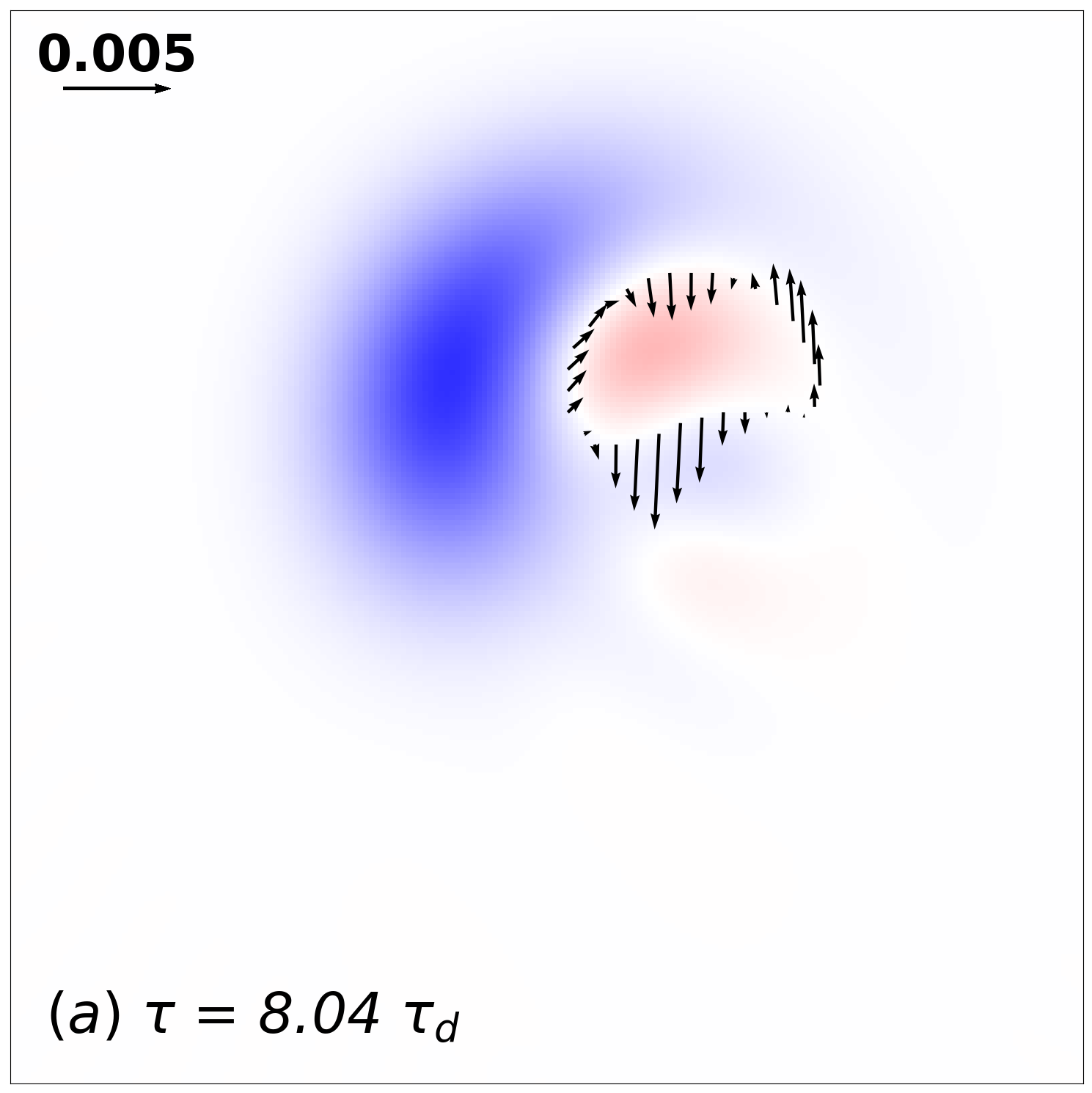}
\includegraphics[width=0.24\textwidth]{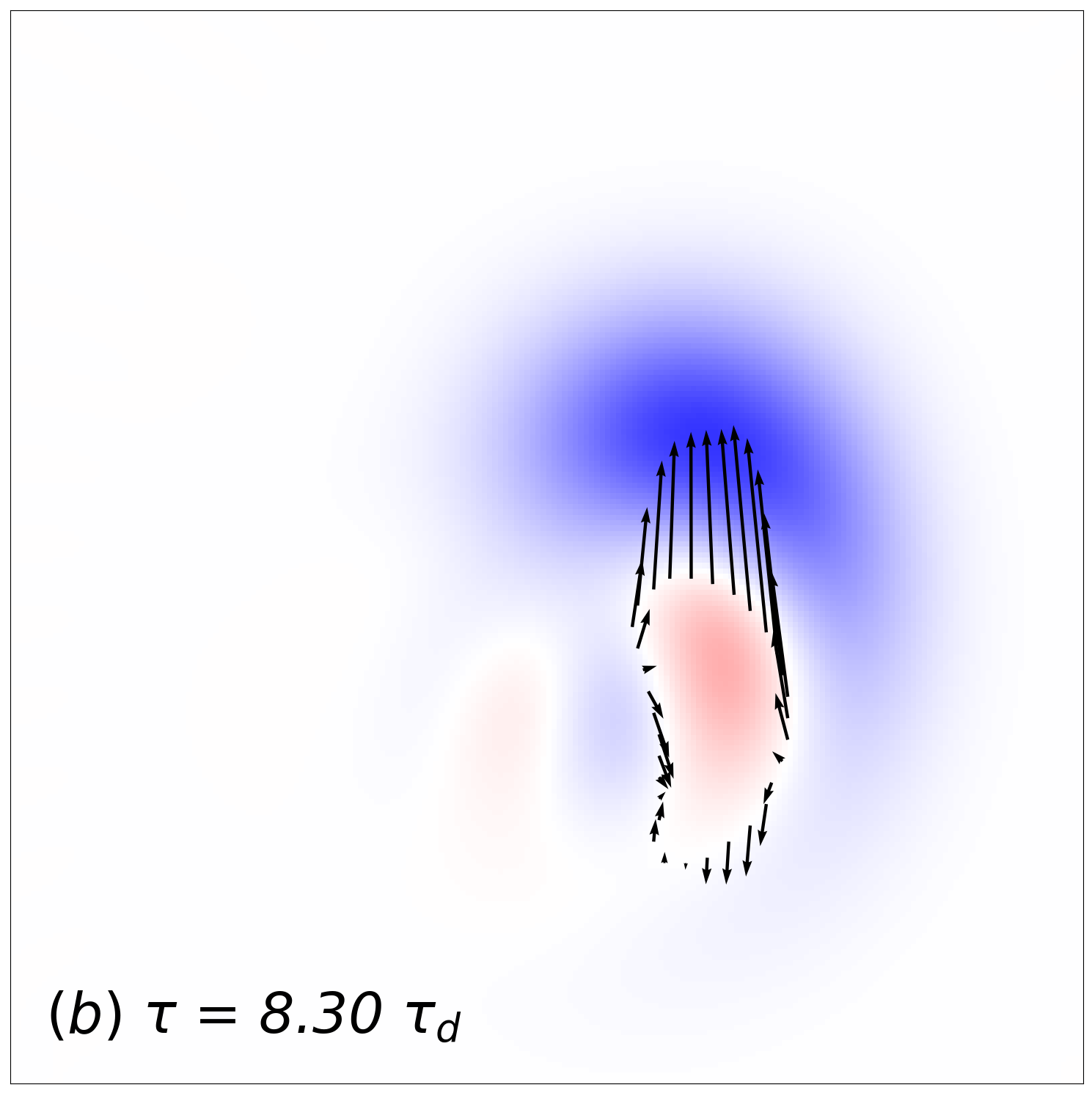}
\includegraphics[width=0.24\textwidth]{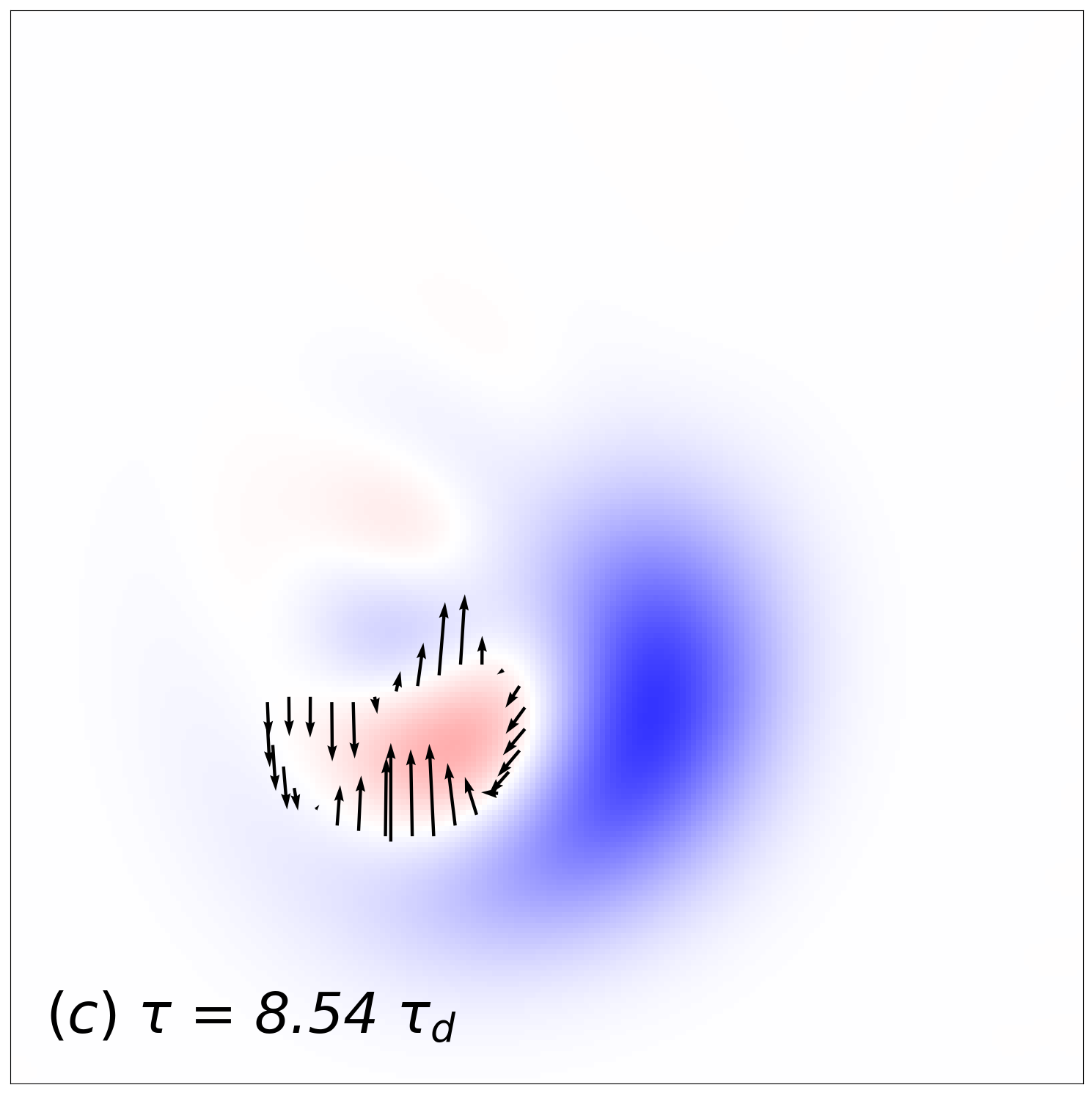}
\includegraphics[width=0.24\textwidth]{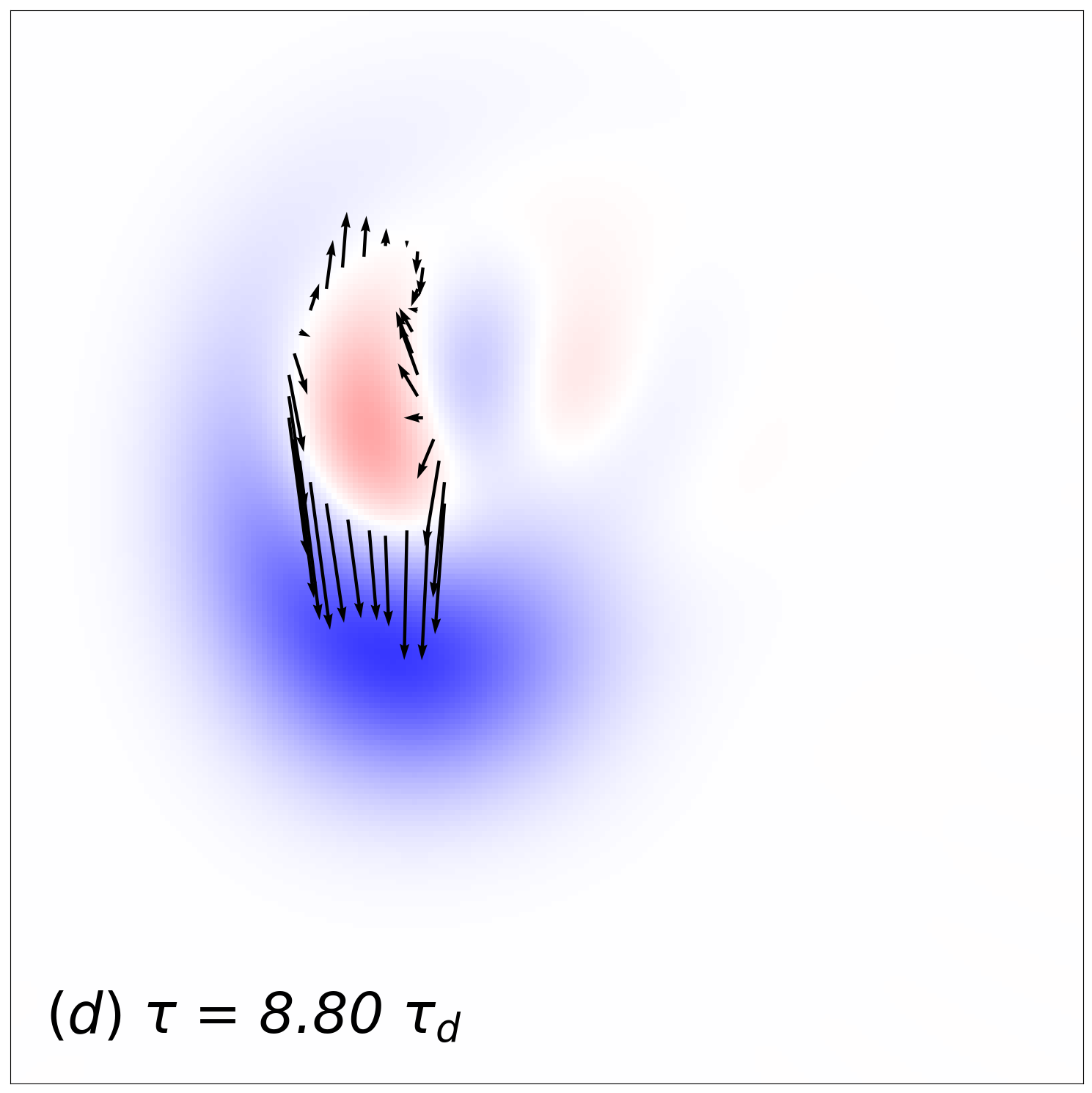}

\includegraphics[width=0.24\textwidth]{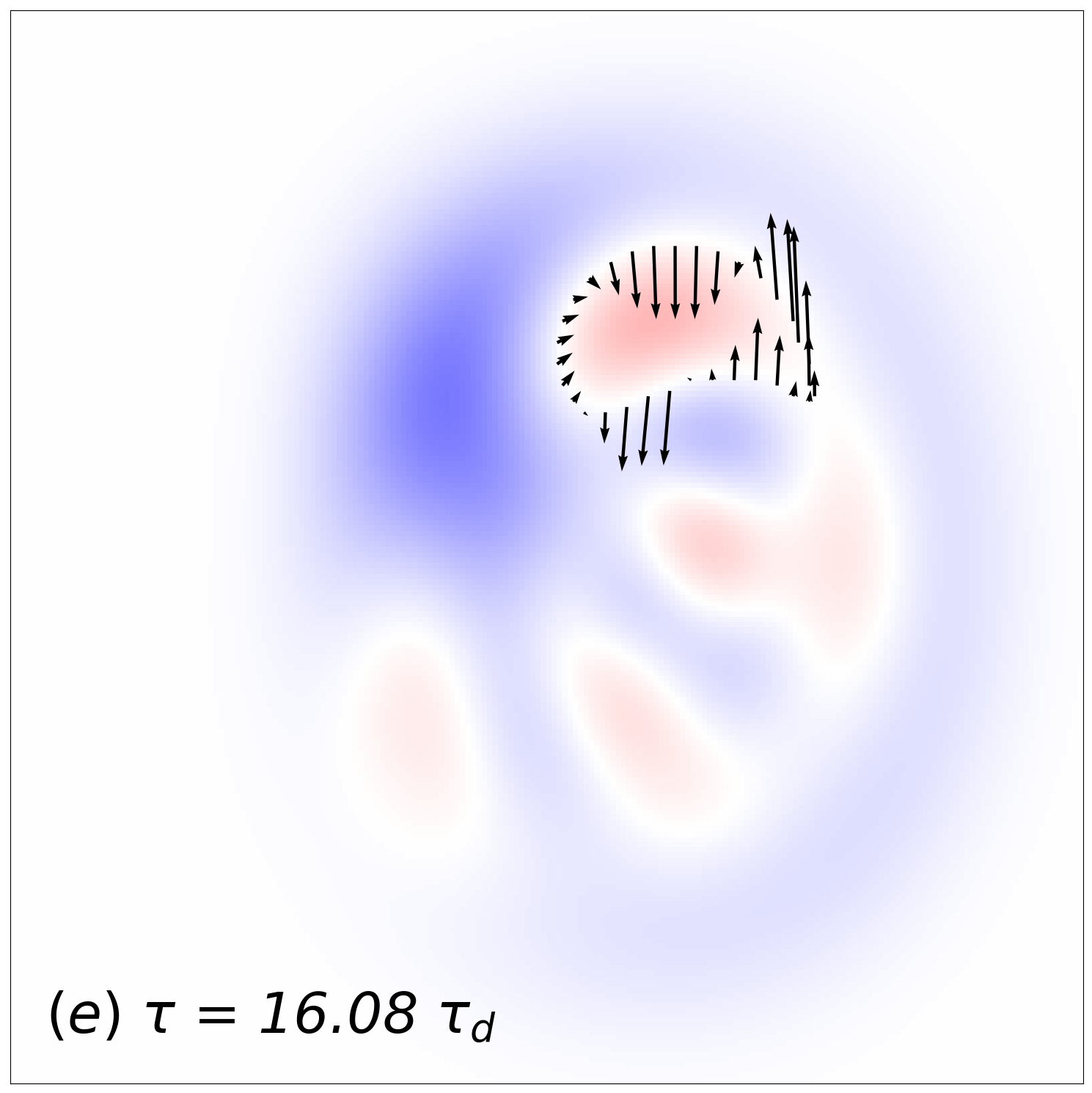}
\includegraphics[width=0.24\textwidth]{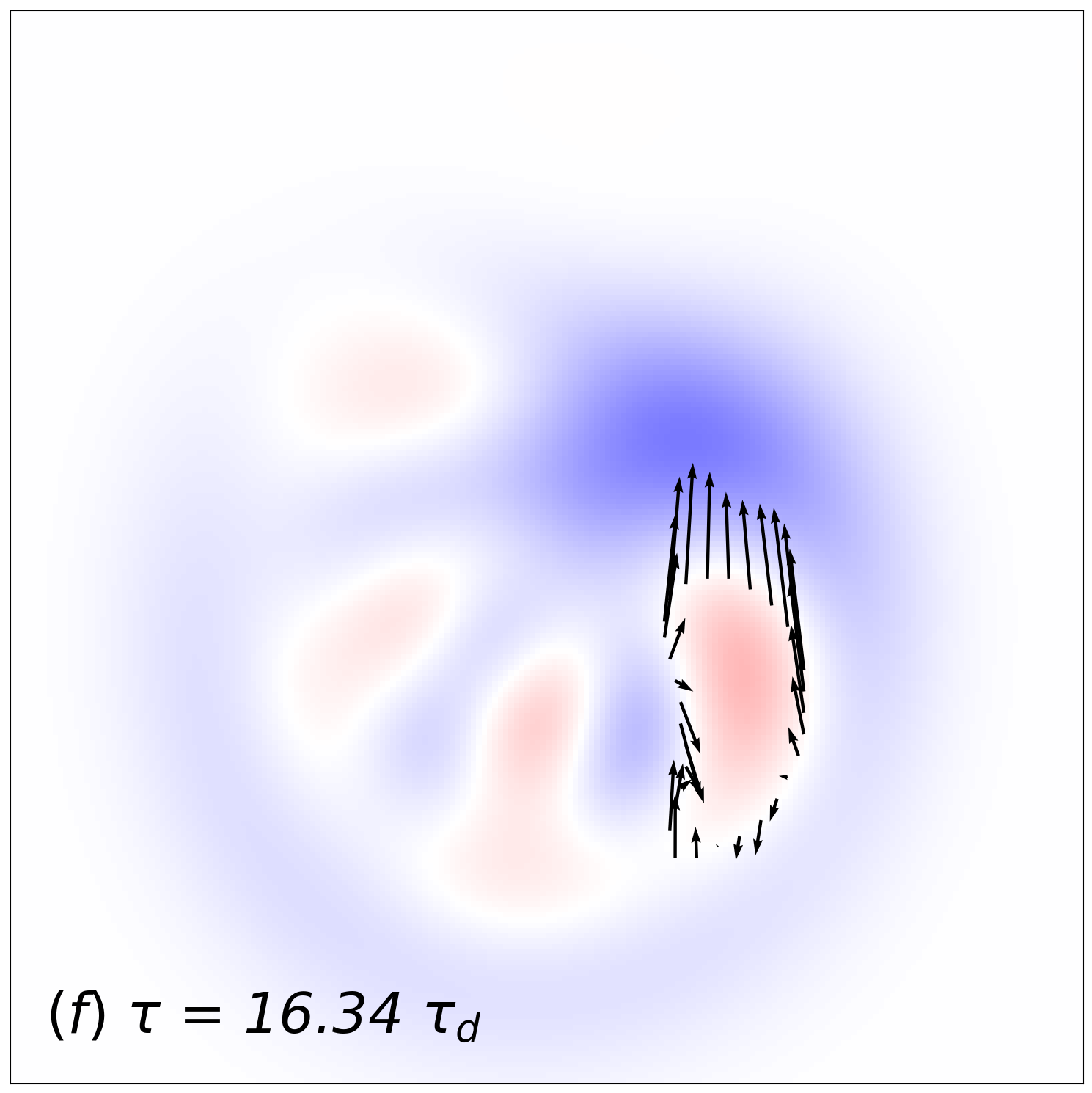}
\includegraphics[width=0.24\textwidth]{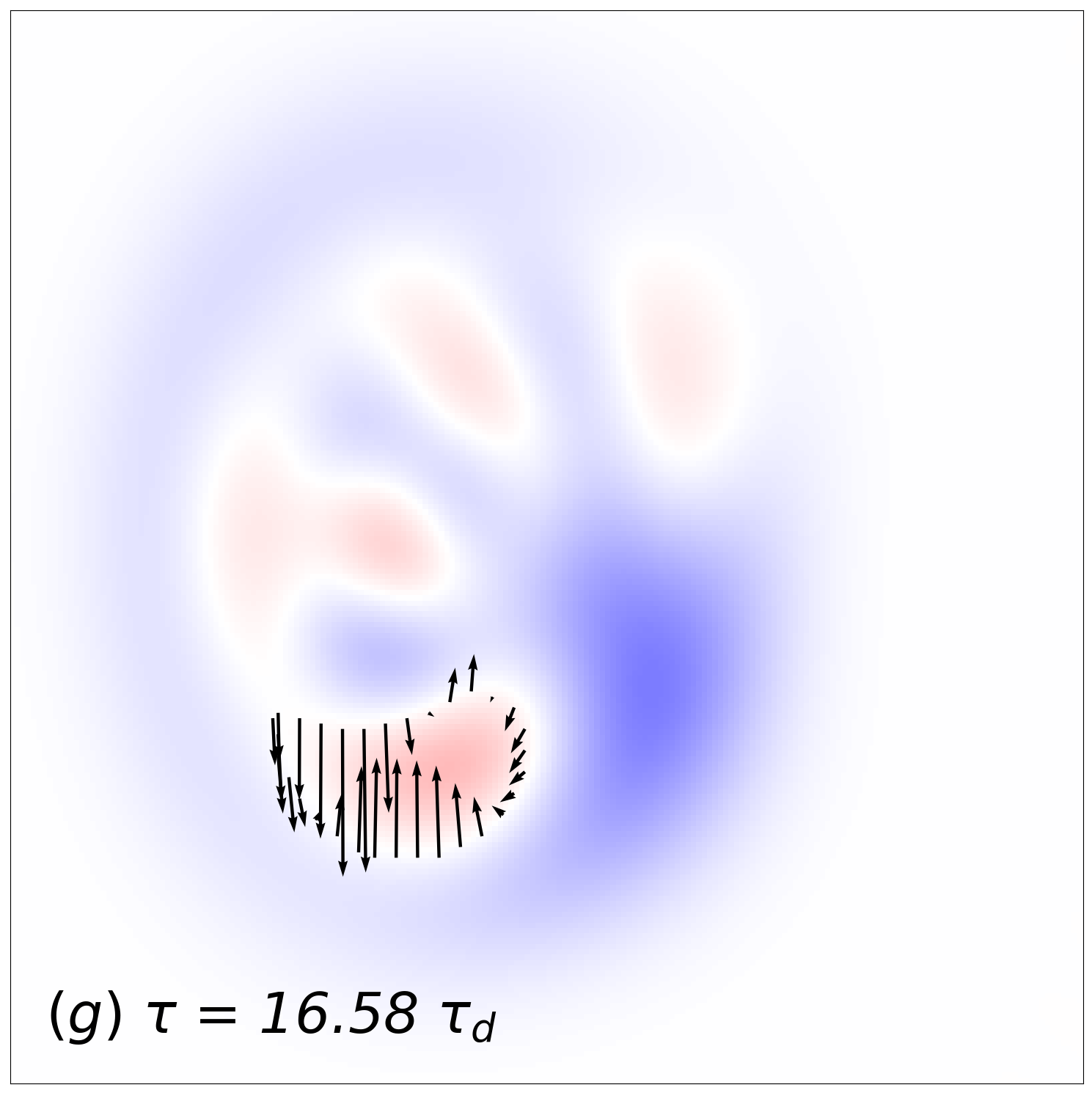}
\includegraphics[width=0.24\textwidth]{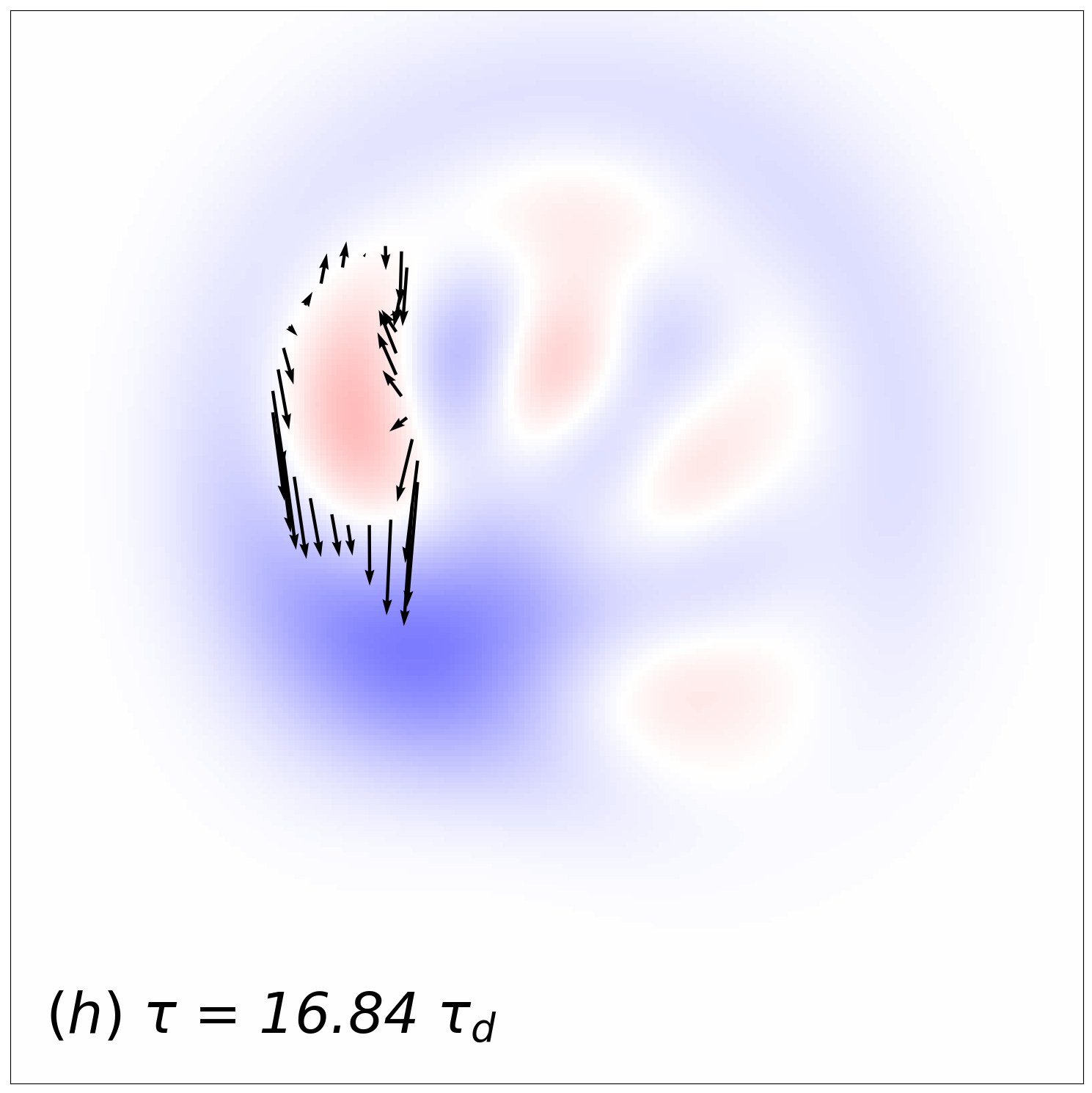}

\includegraphics[width=0.24\textwidth]{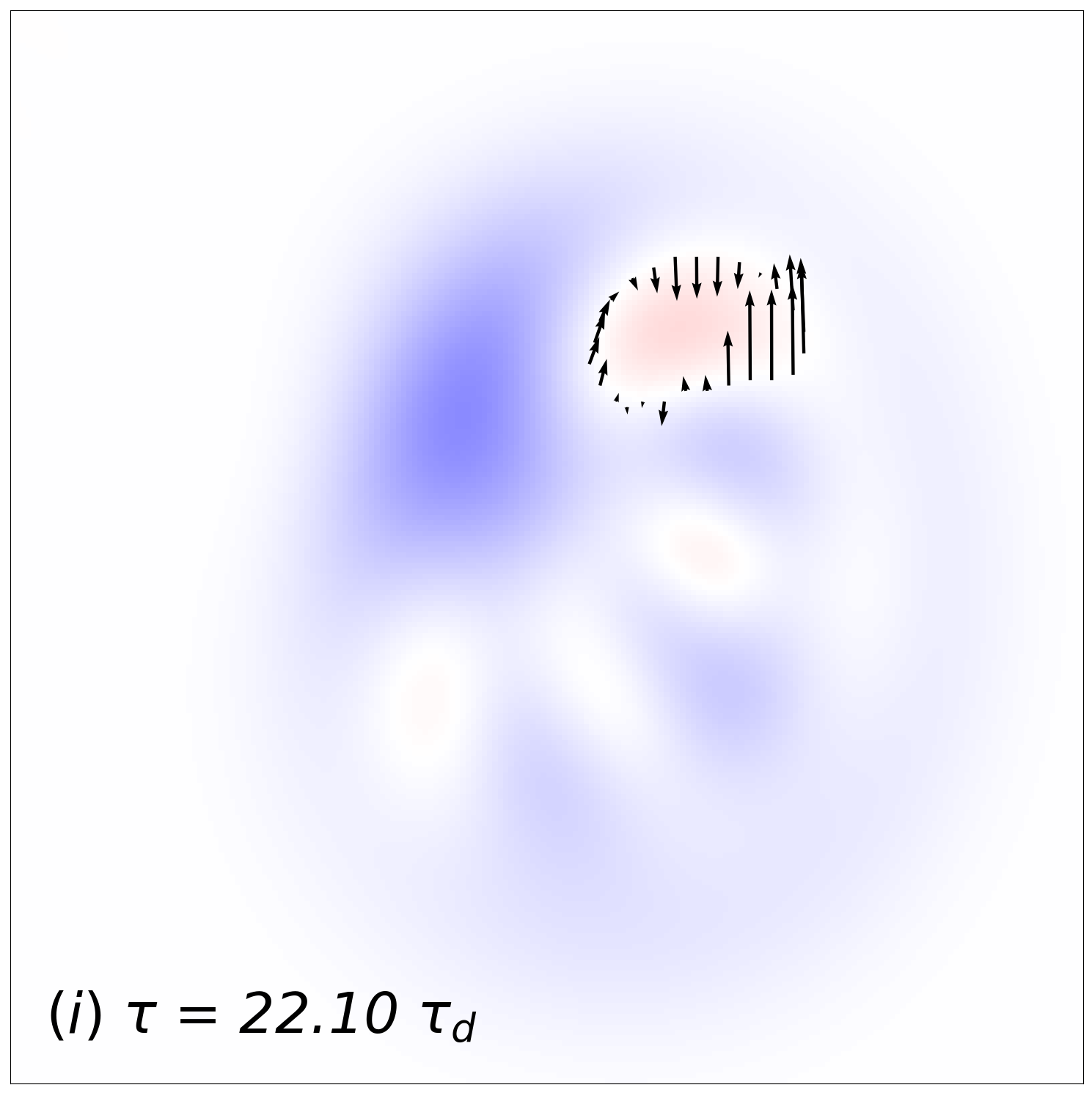}
\includegraphics[width=0.24\textwidth]{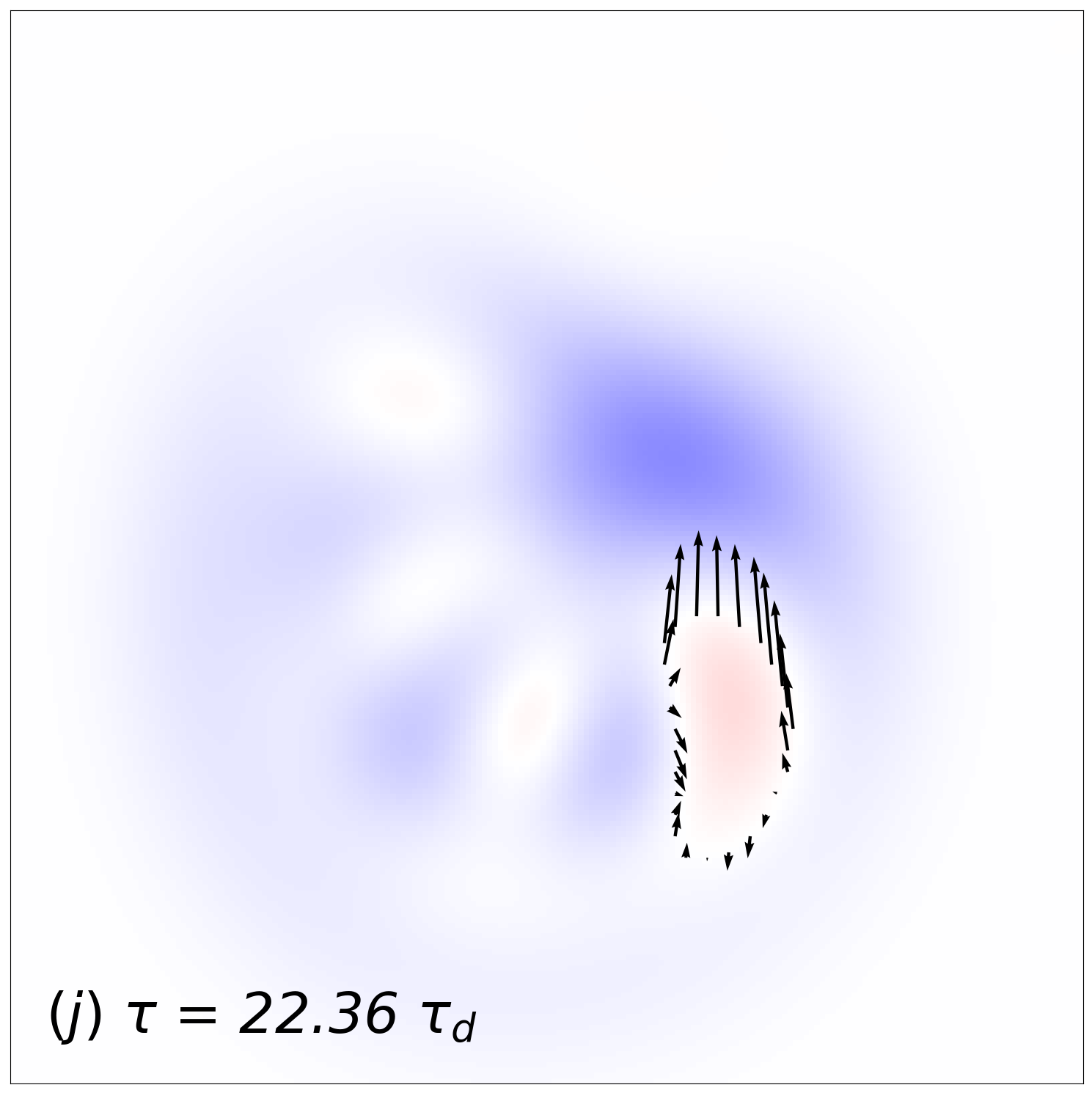}
\includegraphics[width=0.24\textwidth]{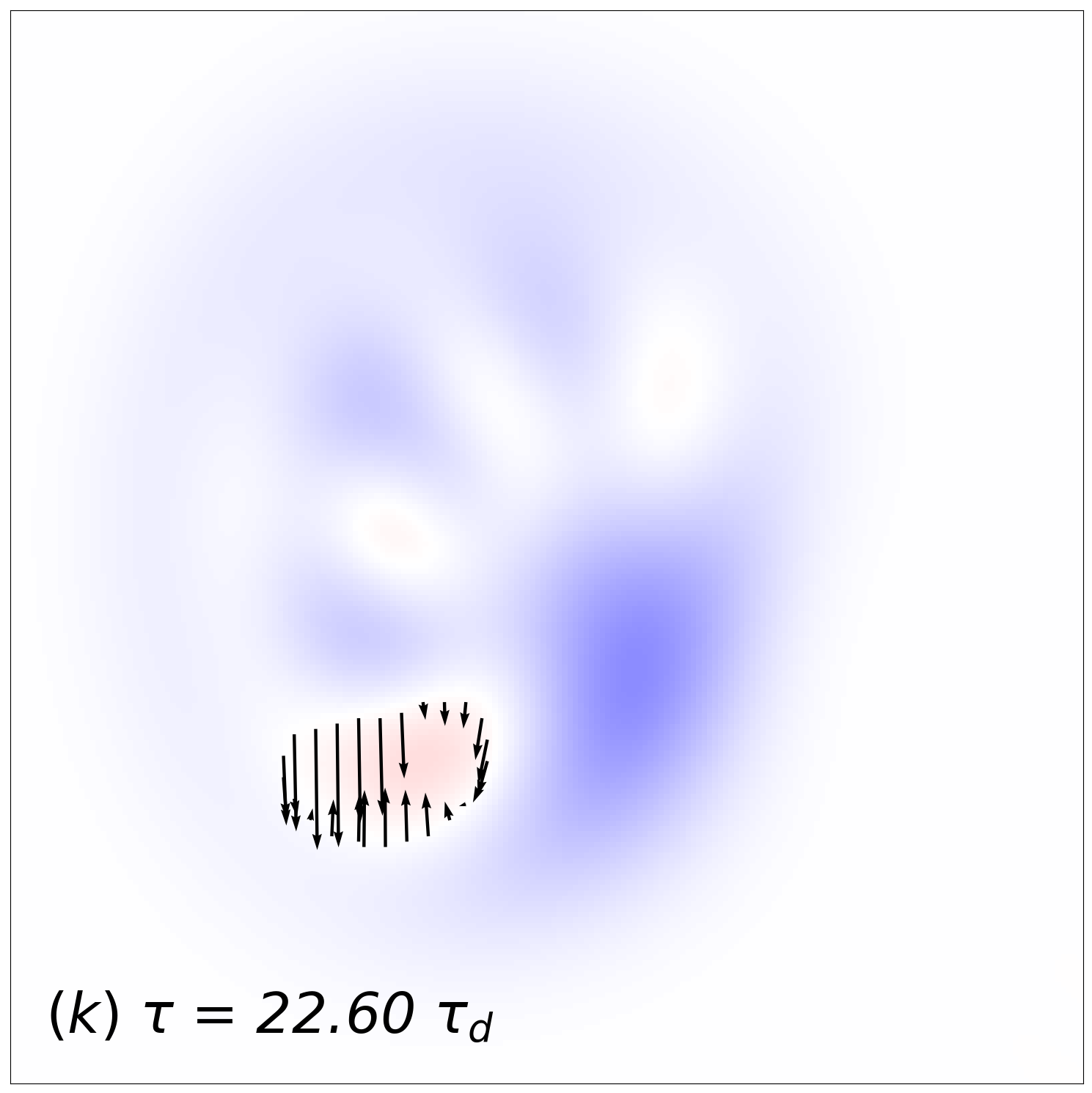}
\includegraphics[width=0.24\textwidth]{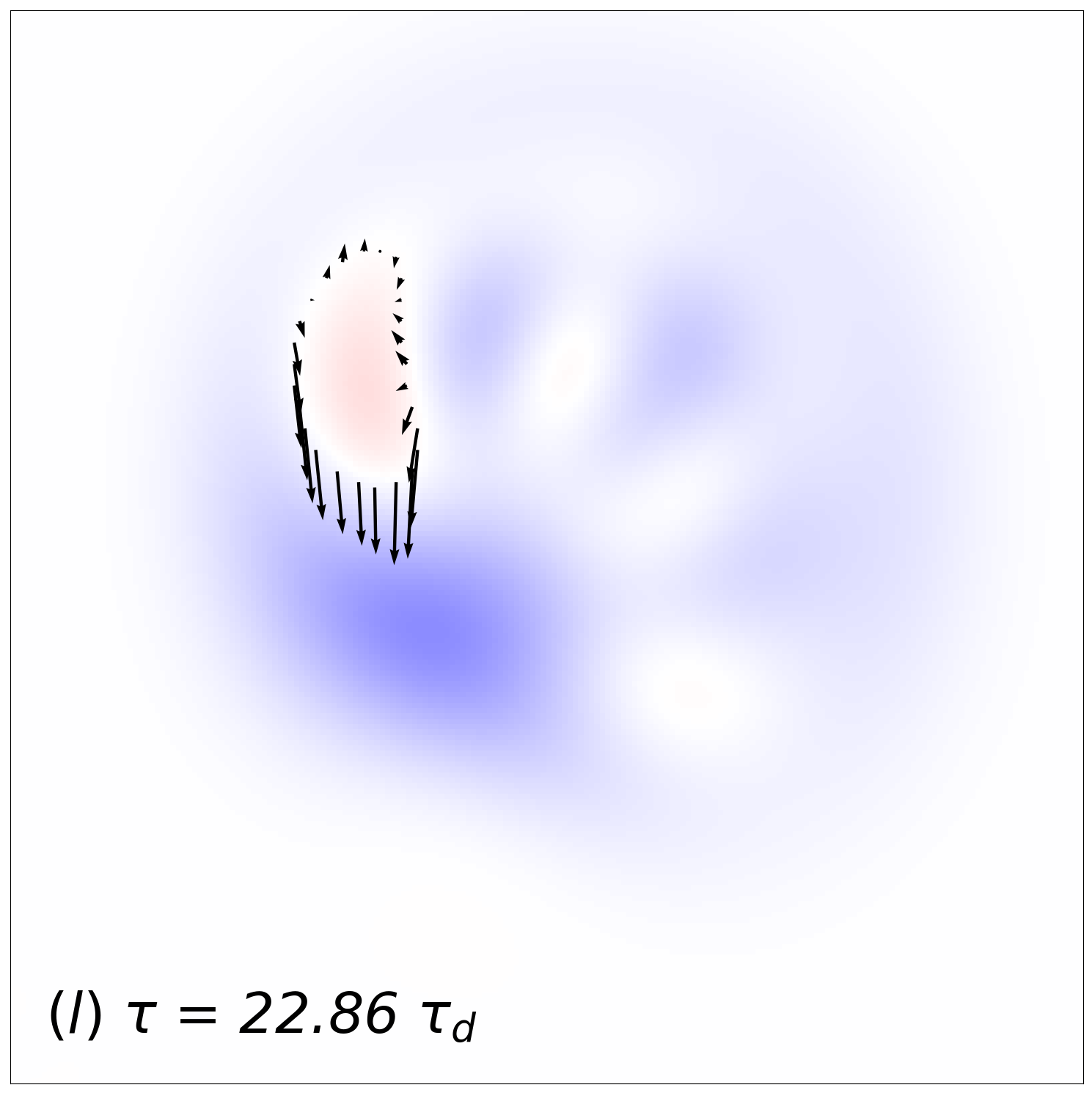}
\caption{Snapshots of evolving Duffing oscillator Wigner function and current vector field ${\mathbf{J}}$ on the boundary of a single negative region for an initial undisplaced coherent state; the damping rate $\gamma=0.01$ and bath temperature $T=0$.}
\label{fig:figure5}
\end{figure*}

\begin{figure*}[htp]
\centering
\includegraphics[width=1\textwidth]{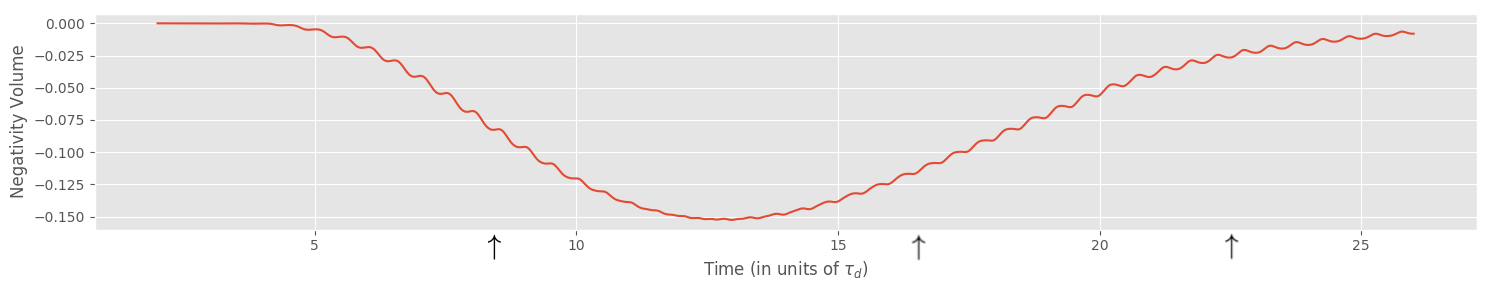}
\caption{Time evolution for the volume of the first sizable negative region to form and the focus of Fig. 5. The arrows indicate approximate times corresponding to the snapshots appearing in the Fig. 5 rows.}
\label{fig:figure6}
\end{figure*}

In particular, for the Duffing oscillator system ($\lambda\neq 0$) on the other hand, we see from Eq.~(\ref{negduffrateeq}) that the rate of change of the region negativity is now governed by two current contributions: the system quantum current $(0, \lambda x\partial_p^2 W/4)$ and the environmental diffusion current~(\ref{dimlessdifffloweq}). Figure \ref{fig:figure5} shows the current flow vector field on the $W=0$ boundary of a particular negative region, comprising the net sum of these two current contributions. Each row of subplots in Fig. \ref{fig:figure5} show the Wigner function at times separated by one quarter of the drive period. The subplots in first row occur at times when the long-term trend is for the negativity volume to grow. Those in the second and third rows occur at times with the opposite long-term behavior.  Figure \ref{fig:figure5}(a) shows a negative region when its volume is slightly increasing due to a net outward boundary current. The region evolves into that which is shown in Fig. \ref{fig:figure5}(b) when the negativity volume is now significantly increasing due to the comparatively larger net outward boundary current. In Fig. \ref{fig:figure5}(c) the region now contracts with the net flow inwards but expands again in Fig. \ref{fig:figure5}(d). This pulsing behavior repeats itself over many periods of the drive, as can be seen in the other plots of Fig. \ref{fig:figure5} and also in Fig. \ref{fig:figure6} which traces the continuous evolution of the same negative region from initial formation to almost complete disappearance; note in particular the small half-drive period oscillations in the negativity volume.

At a given location on the boundary, the environmental diffusion current always flows inwards, thus acting to destroy the negative region. On the other hand, the functional dependence of the system quantum current leads to more varied flow behavior on the boundary and can act to either create or destroy the negative regions; the quantum current must therefore be responsible for the initial generation and possible eventual stabilization of negative regions in the steady state.  

In order to counteract the diffusive inflow, from Eq. (\ref{negduffrateeq}) we necessarily require that
\begin{equation}
\frac{\lambda}{4} \int_{\partial A(t)} ds\, {\mathbf{n}}\cdot\left(0,x\right) \frac{\partial^2 W}{\partial p^2}>0.
\label{negquantumfloweq}
\end{equation}
 Although the orientation and location of the boundary segments along with the magnitude of the term $\partial^2_p W$ can result in quite complicated current flows on the negative region boundaries, there are certain conditions that lead to predictable flow patterns. To aid our intuition, we will focus on the two terms $\mathbf{n}\cdot\left(0,x\right)$ and $\partial^2_p W$ in Eq. (\ref{negquantumfloweq}), which must both be large for there to be a significant quantum current. The $\mathbf{n}\cdot\left(0,x\right)$ term is significant on boundary segments with sizable $x$ coordinates and where the normal vector $\mathbf{n}$ is oriented with a small angle relative to the vertical $p$ axis. At certain times in the evolution over a given drive period, the second derivative $\partial^2_p W$ becomes large at these same boundary segments, resulting in sizable boundary currents inwards or outwards depending on the relative signs of the $\mathbf{n}\cdot\left(0,x\right)$ and $\partial^2_p W$ terms, and the the sign of $\lambda$ which we assume here to be positive. This can be observed whenever the large positive peak is directly above or below the negative region of interest [c.f. Figs. \ref{fig:figure5}(b), (d), (f), (h), (j), and (l)], leading  to large positive $\partial^2_p W$ where $\mathbf{n}\cdot\left(0,x\right)>0$,   and therefore resulting in a  large net outward current on the boundary segment of the negative region that is proximal to the large positive peak. 
 
 In the quadrants where $x$ and $p$ have the same sign [c.f. Figs. \ref{fig:figure5}(a), (c), (e), (g), (i), and (k)], the positive peak is now directly to the left or right of the negative region of interest, leading instead to large positive $\partial^2_x W$,  and having little effect on the $\partial^2_p W$ term. In fact, in these quadrants,  $\partial^2_p W < 0$ on most of the negative region boundary where $\mathbf{n}\cdot\left(0,x\right)>0$, resulting in a net inwards flow.

This change in the direction of the quantum current 
from outwards to inwards on the boundary is the mechanism responsible for the oscillation in the volume of the individual negative region as it cycles clockwise in phase space. Figure \ref{fig:figure6} clearly displays this ``heartbeat" behavior for the single negative region of interest shown in the snapshots of Fig. \ref{fig:figure5}. Arrows on the time axis of Fig. \ref{fig:figure6} indicate the approximate times of the snapshots.

\begin{figure}[htp]
\includegraphics[width=\columnwidth]{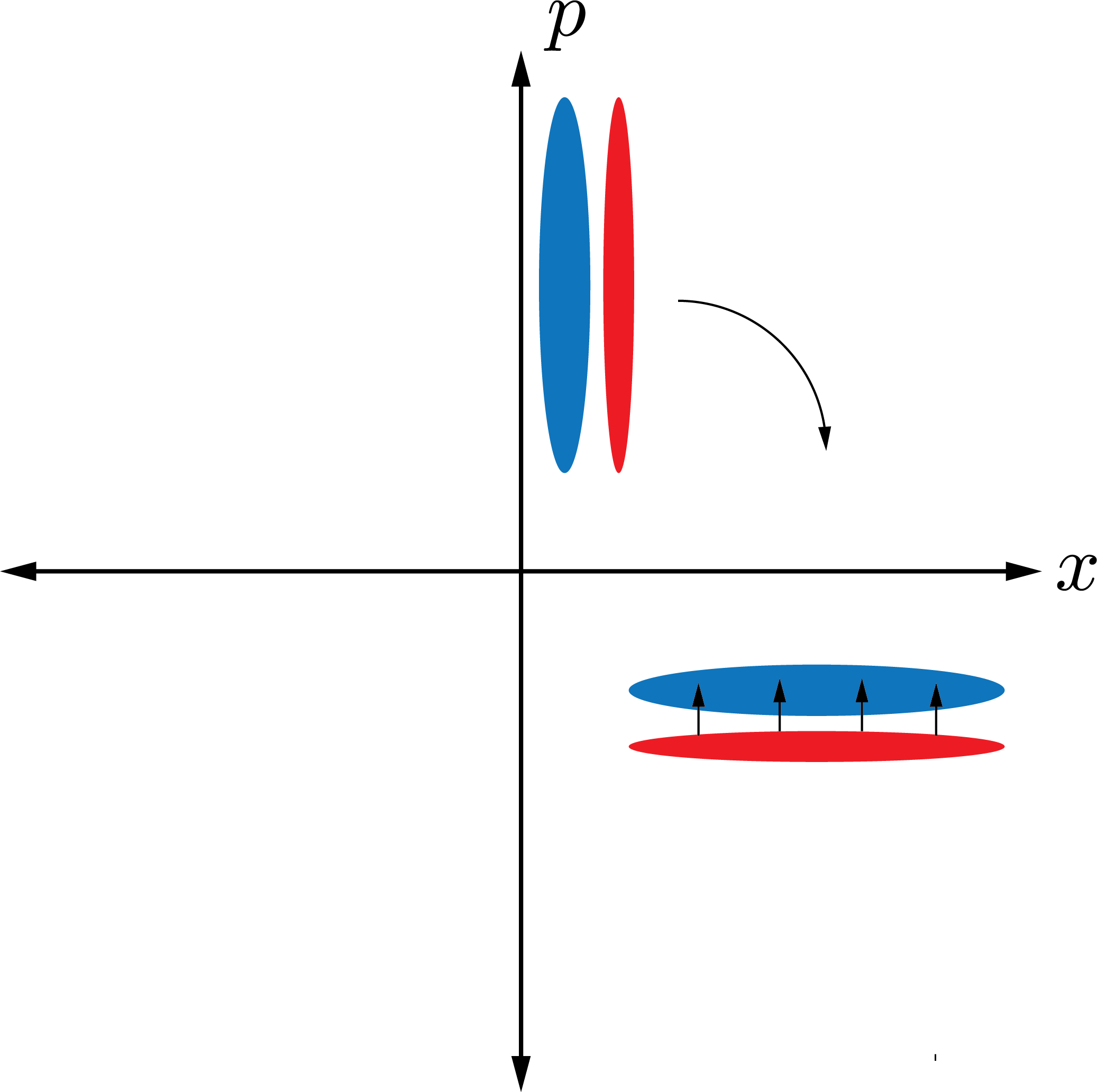}
\caption{Snapshots of an idealized Wigner function distribution geometry that may sustain a significant negative region as it repeatedly cycles clockwise in phase space. Note that the indicated distribution is schematic only and meant to show the relative locations of the positive and negative regions and their overall aspect ratios.  The distribution is squeezed in such a way as to maximize the horizontal dimension when the quantum current can play a dominant role in growing the negative volume, and minimize the horizontal dimension when the net Wigner current would otherwise reduce the negative volume.} 
\label{fig:figure7}
\end{figure}
While significant negative regions develop in the Duffing oscillator numerical example considered above during intermediate times [Fig.~\ref{fig:figure3}(b)], the negative regions practically vanish in the long time limit steady state, even at zero temperature [Fig.~\ref{fig:figure3}(c)]. We have seen that as negative regions transit between adjacent quadrants of phase space, there is an ebb and flow in their size; any stabilization of negativity would be in the sense that significant negativity persists and is repeated at times that are equal modulo the drive period. For this to be possible, the net effect of the quantum current over the drive period must be to increase the negativity volume, thus counteracting the deleterious effects of the diffusion term in Eq.~(\ref{negduffrateeq}). If the negative region does not disappear during one period of the drive $\tau_d$, we can formulate a criterion for this weaker form of stabilization as follows:
\begin{eqnarray}
&&\int_{t_0}^{t_0+\tau_d} dt \bigg[-\frac{\lambda}{4} \int_{\partial A(t)} ds\, {\mathbf{n}}\cdot\left(0,x\right) \frac{\partial^2 W}{\partial p^2} \cr
&&+ \frac{\gamma}{2}\left(n+\frac{1}{2}\right) \int_{\partial A(t)}ds\, {\mathbf{n}}\cdot\nabla W \bigg] \le 0,
\label{negduffperiodeq}
\end{eqnarray}
as $t_0\rightarrow\infty$.

In the above simulations the chosen, example parameter values result in coexisting small and large amplitude stable oscillations for the classical dynamics; the Wigner function must correspondingly spread out through diffusive current flow from its initially narrow and strongly peaked coherent state distribution [Fig.~\ref{fig:figure3}(a)]. As a result, the magnitude of the term $\partial_p^2W$ must decrease overall, and with the small chosen anharmonic coupling strength value $\lambda$~($=0.05$), the system quantum current term  is too weak to counter the deleterious effects of the diffusion term in Eq.~(\ref{negduffrateeq}) and hence be able to stabilize sizable negative regions.

Although we have not yet identified specific examples that lead to the evolution of the Wigner function such that Eq.~(\ref{negduffperiodeq}) is satisfied, there are certain, basic Wigner function distribution geometries that appear to be good candidates. In particular, consider the situation where a negative region precedes a large positive region as they cycle  clockwise in phase space, similar to what is observed in actual simulations as we have discussed above. If, furthermore, the positive region is squeezed such that a large portion of its boundary is proximal to the negative region [c.f. Fig. \ref{fig:figure7}], then according to our above analysis there should be significant growth of the negative region volume in the two quadrants where $x$ and $p$ have opposite sign, while in the other two quadrants where $x$ and $p$ have the same sign there should be comparatively less shrinkage of the negative region volume. Note that the preceeding argument depends on the inflection point $\partial^2_p W=0$  occurring where $W>0$, so that $\partial^2_p W>0$ on the negative region boundary; based on our above numerical simulations of the Duffing oscillator, it is reasonable to assume this over a large portion of the boundary provided that the maximum of the positive peak is relatively large compared to the minimum of the adjacent negative region.

\section{Conclusion}
\label{sec:conclusion}
In this present work, we extended the Wigner phase space formulation of open quantum system dynamics to include a description of the Wigner current vector fields on phase space. This enables the quantum Fokker-Planck equation describing the Wigner function dynamics to be written in the concise form of a continuity equation. The evolving Wigner current was investigated numerically for a harmonic oscillator and a driven Duffing oscillator in the bistable regime, the latter serving as an illustrative anharmonic system.  Through the application of the two-dimensional Gauss's theorem to boundary-enclosed, negative Wigner function regions on system phase space, we saw that the growth and reduction of  negative regions are governed solely by the so-called quantum current due to the system anharmonicity and the diffusion current across the negative region boundaries. By examining the geometric form of these specific contributions to the total Wigner current, we were able to gain some initial insights as to how negative regions might be stabilized, i.e., maintained in the steady state. 

\clearpage
\section*{acknowledgements}
We  thank Paul Nation  for helpful discussions, as well as John Hudson and Susan Schwarz for their assistance with using the Dartmouth Discovery Cluster. W. F. B. and A. J. R. were supported by the National Science Foundation under Grant No. DMR-1807785. W. F. B. also gratefully acknowledges support of a \textit{Gordon F. Hull} Dartmouth graduate fellowship. M. P. B. and O. D. F. were supported by the National Science Foundation under Grant No. DMR-1507383.

\end{document}